\documentclass[prb,amsmath,amssymb, twocolumn, superscriptaddress, aps]{revtex4-2}
\usepackage[english]{babel}
\usepackage{graphicx}
\usepackage{epsfig}
\usepackage{dcolumn}
\usepackage{bm}
\usepackage{rotating}
\usepackage{multirow}
\usepackage{xspace}
\usepackage{dsfont}
\usepackage[colorlinks=true, allcolors=blue]{hyperref}
\usepackage[table]{xcolor}
\usepackage{float}
\usepackage{braket}
\usepackage{soul}
\def\bk{{\bf k}}
\def\bq{{\bf q}}
\def\br{{\bf r}}
\def\bR{{\bf R}}

\def\bfe{{\bf e}}

\def\enk{ {\varepsilon}}

\def\hP{\hat{P}}

\def\hU{\hat{U}}

\def\hpsi{{\hat{\psi}}}
\def\hpsid{{\hat{\psi}^{\dagger}}}

\begin{document}

\title{Origin of phonon decoherence}
\author{Yiming Pan}
\affiliation
{Institute of Theoretical Physics and Astrophysics, Kiel University, 24118 Kiel, Germany}
\author{Christoph Emeis}
\affiliation
{Institute of Theoretical Physics and Astrophysics, Kiel University, 24118 Kiel, Germany}
\author{Stephan Jauernik}
\affiliation
{Institute of Experimental and Applied Physics, Kiel University, 24118 Kiel, Germany}
\author{Michael Bauer}
\affiliation
{Institute of Experimental and Applied Physics, Kiel University, 24118 Kiel, Germany}
\affiliation{Kiel Nano, Surface and Interface Science KiNSIS, Kiel University, 24118 Kiel, Germany}
\author{Fabio Caruso}
\affiliation
{Institute of Theoretical Physics and Astrophysics, Kiel University, 24118 Kiel, Germany}
\affiliation{Kiel Nano, Surface and Interface Science KiNSIS, Kiel University, 24118 Kiel, Germany}
   
\begin{abstract}
Phonon decoherence determines the characteristic timescales over which
coherent lattice vibrations decay, making it a crucial process for
understanding the non-equilibrium dynamics of crystal lattices after
excitation by a pump pulse.  Here, we report a theoretical and computational
investigation of the origin of phonon decoherence within a
first-principles many-body framework.  We derive quantum kinetic equations for
the dynamics of coherent phonons by explicitly accounting for dissipation
processes induced by electron-phonon and phonon-phonon interactions.  The
decoherence rate and frequency renormalization are formulated in terms of the
non-equilibrium phonon self energy, providing a framework amenable for
ab initio calculations.  To validate this approach, we conduct a
first-principles study of phonon decoherence for the elemental
semimetals antimony and bismuth.  The robust agreement with available temperature- and fluence-dependent
experimental data confirms the accuracy of our  theoretical and computational
framework.  More generally, our findings reveal that either electron-phonon
and phonon-phonon coupling can prevail in determining the decoherence time,
depending on the temperature and driving conditions.  Overall, this work fills
a critical gap in the theoretical understanding of phonon decoherence,
providing a predictive framework for determining the timescales of
light-induced structural dynamics in driven solids.
\end{abstract}
	
\maketitle
	
\section{Introduction}
Recent advances in laser technology have expanded the opportunities
to optically control the coherent motion of crystal lattices.
Coherent phonons -- oscillatory  displacements of the atoms from equilibrium -- enable a wide range of non-equilibrium phenomena, including the
initiation of non-thermal phase transitions
\cite{horstmann2020coherent,ZhangWang2019,QiGuanZahn2022,GuanLiuChen2022},
access to metastable phases \cite{LiQiu2019,delaTorre2021,Foerst2011}, control
of material order parameters \cite{GuanChen2023,JuraschekFechner2017}, and the
activation of optically-inactive modes via non-linear phononic principles \cite{JuraschekMeier2020}.
Coherent phonons can undergo a variety of excitation pathways, which depend
on the electronic and crystal structures as well as
driving conditions as light frequency, fluence, and polarization.
Theoretical investigations of coherent phonons have primarily addressed the
excitation principles at the origin of coherent lattice motion
\cite{Scholz1993,Garrett1996,QiShin2009,MERLIN1997207,Juraschek2018,JuraschekFechner2017,Mankowsky2017,Teitelbaum2018,Caruso2023}.
 Conversely, a systematic description of phonon decoherence --  namely, the decay of the coherent state resulting in the progressive disappearance of coherent phonon oscillations -- has  thus far remained uncharted. 

\begin{figure} [b]
\includegraphics[width=0.35\textwidth]{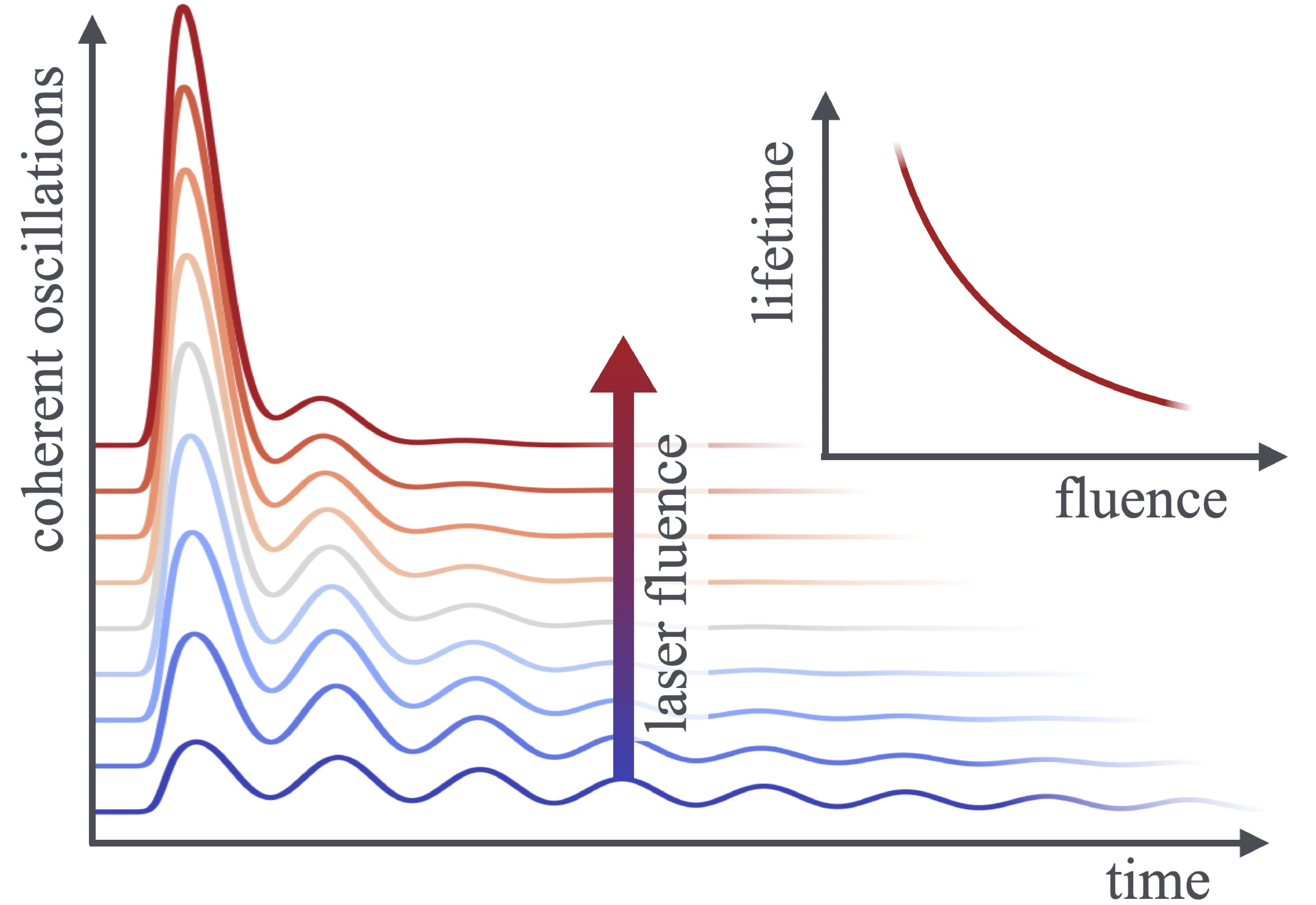}
\caption{Schematic illustration of the characteristic fluence dependence of the coherent phonon amplitude and decoherence. Inset: dependence of the decoherence time on the fluence. }
\label{fig:0}
\end{figure}

Phonon damping has been extensively discussed under equilibrium
condition, that is, in absence of coherent motion \cite{BauerSchmid1998,Cappelluti2006,SaittaLazzeri2008,Park2008,Lazzeri2006,Calandra2010,Giustino2007,caruso_nonadiabatic_2017,NovkoCaruso2020,Marini2024}.  There is a-priori  no formal justification for directly employing the equilibrium theory of phonon damping to model phonon decoherence out of equilibrium. 
Based on the observation that decoherence occurs over timescales
of the order of few picoseconds, its origin is typically attributed
to anharmonic phonon-phonon scattering processes.  The timescales of decoherence 
in semimetals, however, further exhibit a striking dependence of the driving fluence -- namely, the energy per unit area of the pump -- 
suggesting that electron-phonon interactions mediated by photoexcited carriers
can also prevail under suitable conditions~\cite{Cheng2018, Teitelbaum2018}.
This behavior is schematically illustrated in Fig.~\ref{fig:0}: an increase in driving fluence simultaneously enhances the oscillation amplitudes arising from coherent phonons while also accelerating decoherence, resulting in an inverse proportionality between lifetimes and driving fluence (inset).

The description of phonon decoherence has thus far been limited to the inclusion of a
phenomenological damping term into the equation of motion \cite{Scholz1993,JuraschekFechner2017,Juraschek2018}, however, this approach is unsuitable to clarify the origin of decoherence and its dependence on materials properties and excitation conditions. 
Overall, a rigorous and predictive theory 
capable to accounts for the energy dissipation of coherent phonons induced by the interaction with 
other degrees of freedom of the crystal remains elusive.
As decoherence
largely influences the characteristic timescales of coherent lattice motion, this constitutes a critical gap in the understanding of light-induced structural motion in materials,
raising the question of how to systematically account for 
dissipation within ab initio theories and computational frameworks addressing light-induced structural motion.

In this manuscript, we investigate the origin of dissipation in the dynamics
of coherent phonons on the basis of an ab initio many-body approach.  We
derive quantum kinetic equations for the dissipative coherent phonons dynamics
in presence of electron-phonon and phonon-phonon interactions.  Specifically,
the decoherence rates and frequency
renormalization are formulated in terms of the non-equilibrium phonon self
energy due to electron-phonon and phonon-phonon coupling.  We apply this
approach to the elemental semimetals antimony and bismuth, for which a
thorough experimental characterization of  phonon decoherence has been
reported. Overall, our results are in good quantitative agreement with experiments,
and they capture the temperature and fluence dependence of the decoherence. These findings provide  the foundation for future computational studies of phonon decoherence in materials.

\section{Theory of  phonon decoherence} \label{sec:theo}

In this section, we define the theoretical framework for the description of phonon decoherence and frequency renormalization in anharmonic solids. 
The harmonic non-adiabatic Hamiltonian introduced in Ref.~\cite{Stefanucci2023}, and its extension to account for three-phonon scattering processes, constitute the starting point for the description of phonon decoherence in this manuscript. 
Since this is a non-standard Hamiltonian, we discuss its definition in Sec.~\ref{sec:H}. 
In Secs.~\ref{sec:ep} and \ref{sec:pp}, 
we derive explicit expressions suitable for first principles calculations of decoherence rate and frequency renormalization due to electron-phonon and phonon-phonon coupling, respectively.

\subsection{The phonon Hamiltonian}\label{sec:H}

We consider the ab initio nuclear Hamiltonian in presence of electron-nuclei interactions:  
\begin{align}\label{eq:H}
    \hat{H} =\hat{H}_{0} + \hat{H}_{\rm ep} \quad,
\end{align}
where we omitted the Hamiltonian terms that depend solely on
fermionic operators, as they are inconsequential for the 
coherent phonon equation of motions discussed in this manuscript.
Under the assumption of small nuclear displacements, 
the free phonon Hamiltonian $\hat{H}_{0}$ and electron-phonon interaction $\hat{H}_{\rm ep}$ 
can be expanded up to second-order in the displacement amplitude 
$U_{{\bf q}\nu}$ \cite{Stefanucci2023}:
\begin{align}
    \hat{H}_{0} &= \frac{1}{2}\sum_{\bq\nu} \hP_{-\bq\nu}\hP_{\bq\nu} + \frac{1}{2}\sum_{\bq\nu\nu'}\hU_{-\bq\nu'}K_{\bq\nu\nu'}\hU_{\bq\nu} \label{eq:H0} \quad,\\
    \hat{H}_{ep} &= \int  d\br [\hat{n}(\br)-n_0(\br) ]\sum_{\bq\nu} g_{\bq\nu}(\br) \hU_{\bq\nu} \label{eq:Hep} \quad.
\end{align}
The density operator $\hat{n}(\br) =
\hpsid(\br)\hpsi(\br) $
is given in terms of fermionic 
field operators
$\hpsi(\br)$ and $\hpsi^\dagger(\br)$.
 The electron-phonon coupling function is defined as:
\begin{align}
g_{\bq\nu}(\br) = 
\sum_{p \kappa} \frac{1}{\sqrt{NM_\kappa}} e^{i \bq \cdot \bR_p } 
\bfe^{\kappa}_{\bq \nu} \cdot \frac{\partial V^{\rm ion}(\br)}{\partial \boldsymbol{\Delta \tau}_{\kappa p}}\quad.
\end{align}
Here, ${\bf R}_p$ is a crystal lattice vector, $M_\kappa$ the mass of the $\kappa$-th nucleus, $\bfe^\kappa_{\bf q\nu}$ denotes a normal-mode basis vector, 
${\boldsymbol  {\Delta \tau}}_{\kappa p}$ is the 
displacement vector of the $\kappa$-th atom in the $p$-th unit cell 
from its equilibrium coordinate, and 
 $ V^{\rm ion} = - 
\sum_{\kappa p} {Z_{\kappa}}{| \br - {\boldsymbol \tau}_{\kappa p}  |^{-1} } $ is the electron-nuclei potential.  
The sum over $p$ and $\kappa$ runs over all unit cells in the Born-von-Karman supercell and over all atoms in the unit cell, respectively. 

The operator $\hU_{\bq\nu}$ has units of a length and characterizes the coherent displacement of the lattice along the mode
$(\bq, \nu)$. It is related to the Cartesian representation of 
the displacement operator ${\boldsymbol  {\Delta \hat\tau}}_{\kappa p}$  via the 
normal coordinate transformation: 
\begin{align}
{\boldsymbol {\Delta \hat\tau}}_{\kappa p} = (NM_\kappa)^{-\frac{1}{2}}
\sum_{\bf q \nu} 
 e^{i \bq \cdot \bR_p } {\bf e}^\kappa_{\bf q\nu} \hat{U}_{\bf q\nu}\quad.
\end{align}
$\hP_{\bq\nu}$ is the canonical momentum defiend by the 
commutation relation $[\hU_{\bq\nu}, \hP_{\bq'\nu'}]  =
i\hbar\delta_{-\bq,\bq'}\delta_{\nu\nu'}$. In Eq.~\eqref{eq:H0},
$K_{\bq\nu\nu'}$ is the normal coordinate representation of the 
elastic tensor, defined as: 
\begin{align}
K_{\bq\nu\nu'} = \sum_{\kappa\kappa' p } 
e^{i\bq\cdot \bR_p}{\bf e}^\kappa_{\bf q\nu} \cdot
\frac{{\bf K}_{\kappa p , \kappa' p'}}{\sqrt{M_\kappa M_{\kappa'}}}
\cdot {\bf e}^{\kappa'}_{\bf q\nu'}\quad,
\end{align}
where 
$
{\bf K}_{\kappa p\kappa'p'} = 
{\boldsymbol \Delta}_{ \substack{\kappa p \\ \kappa 'p ' }}
 E_{n-n}
+\int d \mathbf{x}
n^0(\mathbf{x}) 
{\boldsymbol \Delta}_{ \substack{\kappa p \\ \kappa 'p '} }
V^{\rm ion}, 
$
 we introduced the abbreviation 
${\boldsymbol \Delta}_{\substack{\kappa p \\ \kappa 'p '} } 
= \frac{\partial^2 }{
\partial {{\boldsymbol {\Delta \tau}}_{\kappa p} } 
 \partial {{\boldsymbol {\Delta \tau}}_{\kappa' p'} }}$
, 
and 
$ E_{n-n}  = \frac{1}{2} \sum_{\kappa p \neq \kappa'p'} 
{Z_\kappa Z_\kappa'} {|{\boldsymbol \tau}_{\kappa p} - 
{\boldsymbol \tau}_{\kappa' p'}|^{-1} }
$ is the nuclei-nuclei Coulomb energy. 

The relation between the Hamiltonian defined by Eq.~\eqref{eq:H0} 
and the adiabatic (Born-Oppenheimer) harmonic Hamiltonian -- the usual starting point for the definition of phonons in solids --  has been discussed in detail in Ref.~\cite{Stefanucci2023}. 
The key differences of relevance for this work are briefly summarized below: 
 (i) Eq.~\eqref{eq:H0} does not rely on the Born-Oppenheimer approximation; (ii)
 the elastic tensor $K_{\bq\nu\nu'}$ 
 only includes the contribution of the bare nuclei 
 and it is in general not positive-definite \cite{van_leeuwen2004,Stefanucci2023}. 
 As a consequence, the eigenvalues of ${\bf K}$ are not positive definite, and cannot be interpreted as normal vibrational frequencies of the lattice. 
(iii) Adiabatic-harmonic phonon frequencies are only defined in terms of the adiabatic phonon self-energy;
(iv) the lattice dynamics depends on the time-dependent electron density $\hat{n}(\br, t)$ via Eq.~\eqref{eq:Hep}. 

\subsection{Decoherence and frequency renormalization mediated by electron-phonon coupling}\label{sec:ep}

In the following, we proceed to derive the 
damped equation of motion of the nuclear displacements 
$U_{\bq\nu}$. 
Heisenberg picture is implied throughout. 
We begin by considering the Heisenberg equation of motion for 
displacement operator: 
\begin{align}\label{eq:HEOM}
\frac{\partial ^2  U_{\bq\nu}}{\partial t^2}  = 
-\hbar^{-2} \langle [\hat U_{\bq\nu},[\hat U_{\bq\nu},\hat H]] \rangle\quad. 
\end{align}
Making use of Eqs.~\eqref{eq:H}-\eqref{eq:Hep} alongside with the
commutation relations for $U_{\bq\nu}$ and $P_{\bq\nu}$ one readily
obtains:
\begin{align}\label{eq:coh1}
\frac{\partial^2 U_{\bq\nu}}{\partial t^2} + \sum_{\nu'}K_{\bq \nu'
\nu}U_{\bq\nu^{\prime}} = -\int d\br g_{-\bq\nu}(\br) \delta n(\br,
t)  \quad.
\end{align}
The right-hand side represents the force exerted on the
lattice by a time-dependent density fluctuation $\delta n(\br, t)=n(\br,
t)-n_0(\br) $, where $n_0(\br) $ denotes the equilibrium density. 
Here, we consider the case in which
finite values for $\delta n(\br, t)$ result from
time-dependent perturbations, such as, e.g., an external driving field coupled to the electron density, 
resulting in the excitation of coherent phonons. 

Equation~\eqref{eq:coh1} resembles the equation of motion of a undamped driven harmonic oscillator, which is reflected by the absence of a term proportional to $\partial U_{\bq\nu}/\partial t$. 
However, since Eq.~\eqref{eq:coh1} is an exact result for the Hamiltonian
specified by Eqs.~\eqref{eq:H}-\eqref{eq:Hep}, we expect the effects
of decoherence to be encoded in the density fluctuation $\delta n(\br,
t)$. 
To retrieve an explicit expression for the decoherence rate mediated by the electron-phonon coupling, 
we express  $\delta n(\br,
t)$ making use of linear response theory: 
\begin{align}\label{LR}
    \delta n(\br, t) = \delta &n(\br, t_0) \\
&+ \int d{\br'} \int_{t_0}^t d{t'} \chi^{r}(\br, t ; \br', t')\delta V^{\rm ion}(\br', t') \nonumber \quad,
\end{align}
where $\delta n(\br, t_0)$ is the variation of electronic density induced by
the time-dependent perturbation (i.e., a pump pulse), 
responsible for triggering the coherent dynamics of the lattice, 
and the second term accounts for density fluctuations induced by 
the electron-phonon interactions. 
$\chi^{r}(\br, t ; \br', t')$ is the retarded
density-density response function, and 
$\delta V^{\rm ion}(\br,t) = \sum_{\bq\nu} g_{\bq \nu} (\br) U_{\bq\nu}(t)$.

By introducing the adiabatic and nonadiabatic
density responses to the electron-phonon interactions
-- $\delta n^{\rm A}$ and $\delta n^{\rm NA}$, respectively --,
we can rewrite Eq.~\eqref{LR} as:
\begin{align}\label{LR1}
    \delta n(\br, t) = 
\delta &n(\br, t_0) 
+ \delta n^{\rm A}(\br, t) 
+ \delta n^{\rm NA}(\br, t) \quad.
\end{align}
The adiabatic density response 
arises from the instantaneous response of the density \cite{Baroni2001,Giannozzi1991}  to
the nuclear displacement $U_{\bq\nu}$, and 
it can be derived by assuming a parametric dependence in the form 
 $\delta n^{\rm A}(\br, t)
= \delta n^{\rm A} (\br ; U_{\bq\nu}(t))$, leading to ({\it cf}. Appendix B or Ref.~\cite{Stefanucci2023}): 
\begin{align}\label{eq:nA}
\delta n^{\rm A} (\br, t) 
&= \int  d{\br}'\chi^{r}(\br,\br' ;  0) \sum_{\nu'}g_{\bq\nu'}(\br') U_{\bq\nu'}(t)  \quad,
\end{align}
where $\chi^{r}(\br,\br' ;  0)$  is the Fourier transform of 
$\chi^{r}(\br,t ; \br', t' )$ at zero frequency.  

Substitution of Eqs.~\eqref{LR}-\eqref{eq:nA} into \eqref{eq:coh1} enables after   
few algebraic manipulations to rewrite the equation of motion for 
coherent phonons in presence of electron-phonon interactions 
in the form~\cite{SI}: 
\begin{align}\label{eom_Q_eph}
\frac{\partial ^2 U_{\bq\nu}}{\partial t^2} + 2\Gamma^{\rm ep}_{\bq\nu} \frac{\partial U_{\bq\nu}}{\partial t}
&+\Omega_{\bq \nu}^2U_{\bq\nu}  \\
&=\nonumber
-\int d\br g_{-\bq\nu}(\br)\delta n(\br, t_0) \quad.
\end{align}
This expression resembles the equation of motion for a 
driven-damped harmonic oscillator, with the general solution $U_{\bq\nu}(t) = U_{\bq\nu}^0 e^{-\Gamma^{\rm ep}_{\bq\nu}t} {\rm sin} (\Omega_{\bq\nu}t+\varphi)$. 
The quantities $ \Gamma^{\rm ep}_{\bq\nu}$ and 
$\Omega_{\bq\nu}$ can be identified with the decoherence rate and with the renormalized vibrational frequency, respectively, and 
are defined as: 
\begin{align}\label{eq:PiNA}
  \Gamma^{\rm ep}_{\bq\nu} &= - {\rm Im}\Pi_{\bq\nu}^{\rm ep, NA}  \quad,\\ 
 \Omega_{\bq\nu}^2 &= \omega_{\bq\nu}^2 + {2\omega_{\bq\nu}\rm Re}  \Pi_{\bq\nu}^{\rm ep, NA} \label{eq:Omega}\quad.
\end{align} 
Here, $\Pi_{\bq\nu}^{\rm ep, NA}$ is  the non-adiabatic phonon self energy 
due to the electron-phonon interaction:
\begin{align}\label{phself_e}
    \Pi^{\rm ep, NA}_{\bq\nu} = &\frac{1}{2\omega_{\bq\nu}} \int  d\br d\nonumber
\br' g_{-\bq\nu}(\br)\chi^r(\br,\br' ;\omega_{\bq\nu})g_{\bq\nu}(\br')  \\ 
-
&\frac{1}{2\omega_{\bq\nu}} \int d\br d\br'
g_{-\bq\nu}(\br)\chi^r(\br,\br' ;0)g_{\bq\nu}(\br')   \quad.
\end{align}
In Eq.~\eqref{eq:Omega} we introduced the Born-Oppenheimer (harmonic) phonon frequencies $\omega_{\bq\nu}$, 
which are related to the elastic tensor $K_{\bq\nu\nu'}$ via: 
\begin{align}
 {\omega}^2_{\bq\nu} &= K_{\bq\nu}\delta_{\nu\nu'} + {2\omega_{\bq\nu}\rm Re}  \Pi_{\bq\nu}^{\rm ep, A} \quad,
\end{align}
where the adiabatic phonon self energy  $\Pi_{\bq\nu}^{\rm ep,A}$ is defined as:
\begin{align}
 \Pi_{\bq\nu}^{\rm ep,A} =  \frac{1}{2\omega_{\bq\nu}}\int d{\br} d{\br}' g_{-\bq\nu}(\br)\chi^{r}(\br,\br' ; 0) g_{\bq\nu}(\br')   \quad.
\end{align}

Equations~\eqref{eom_Q_eph}-\eqref{phself_e} are the central result of our work, and  
their detailed derivation is reported in the Supplemental Material (SM) \cite{SI}. 
These expressions formally relate the decoherence rate of coherent phonons $\Gamma_{\bq\nu}$
and their frequency renormalization with the non-adiabatic phonon self-energy, thus justifying the approximate identification of coherent phonon lifetimes  with the ordinary phonon lifetimes in the vicinity of equilibrium. 
Additionally, the close relation of Eq.~\eqref{phself_e} with the non-adiabatic 
phonon self-energy derived within the equilibrium 
Green's function formalism enables the application of the 
available implementations for the phonon self-energy to study decoherence and 
softening of coherent phonons. 
In particular, the phonon self-energy introduced above coincides with the expressions reported in 
 non-equilibrium \cite{Stefanucci2023} 
Green's function formulation of the electron-phonon problem, 
which reduce to the equilibrium Green's function formalism \cite{Giustino_RMP2017} for electronic occupations 
described by Fermi-Dirac statistics.
We note that in Eqs.~\eqref{phself_e} the self-energy is expressed in terms 
of the  reducible function and in terms of the bare electron-phonon vertex $g$. 
Alternatively, the self-energy can be recast in terms of the 
irreducible response function, involving a bare
electron-phonon  coupling matrix element 
$g_{\bq\nu}(\br)$ and a dynamically screened coupling matrix elements 
\cite{BergesGirotto2023,Giustino_RMP2017,Stefanucci2023,Marini2023,Caldarelli2024,StefanucciPerfetto2025}.

\subsection{Decoherence and frequency renormalization mediated by phonon-phonon coupling}\label{sec:pp}

In the following, we extend the formalism developed in Sec.~\ref{sec:ep} to describe the effects of phonon-phonon interactions on the 
decoherence and renormalization of coherent phonons. 
Phonon-phonon scattering underpins the temperature dependence of the  phonon decoherence, and it constitutes the leading dissipation mechanisms 
in crystals with low free-carrier concentrations (semiconductors, insulators, and semimetals) 
\cite{Sun2021,Sayers2023,Trovatello2020,Jeong2016}.
We hereby restrict ourselves to consider third-order anharmonicities in the Born-Oppenheimer approximation, neglecting fourth- and higher-order terms and non-adiabatic effects.  
Under these approximations, the phonon-phonon Hamiltonian can be expressed as:
\begin{equation}
    \hat{H}_{\rm anh} = \frac{1}{3!}\sum_{\bq\nu\bq'\nu'\bq''\nu''} V^{(3)}_{\bq\nu,\bq'\nu',\bq''\nu''} \hU_{\bq\nu}\hU_{\bq'\nu'}\hU_{\bq''\nu''}\quad.
\end{equation}
$V^{(3)}_{\bq\nu,\bq'\nu',\bq''\nu''} $ denotes the phonon-phonon coupling matrix elements, which involves the third-order derivative of Born-Oppenheimer potential energy surface relative to the displacements $U_{\bq\nu}$. It can be estimated from 
first principles via finite-difference calculations based on density-functional theory \cite{LI20141747}. 
The translational invariance of the Hamiltonian requires crystal momentum conservation, $\bq + \bq'+\bq'' = {\bf G}$, where $\bf G$ is a reciprocal lattice vector. To focus on phonon-phonon coupling, we omit terms arising from the electron-phonon interactions, and we note that all corrections are additive at linear order.  
The coherent-phonon equation of motion of an anharmonic crystal is thus given by:
\begin{align}\label{eom_phph}
     \frac{\partial^2 U_{\bq\nu}}{\partial t^2} +& \omega_{\bq\nu}^2U_{\bq\nu} = 
     \nonumber \\ - &\frac{1}{2} \sum_{\bq'\nu'\bq''\nu''}  V^{(3) }_{-\bq\nu,\bq'\nu',\bq''\nu''} \langle \hU_{\bq'\nu'}\hU_{\bq''\nu''} \rangle\quad .
\end{align}
Similarly to the case discussed in Sec.~\ref{sec:ep}, Eq.~\eqref{eom_phph} lacks a term 
proportional to $\partial U_{\bq\nu} / \partial t$, which may be promptly identified with the source of 
decoherence due to anharmonicities. 
In this case, decoherence is expected to arise from the two-phonon density matrix 
$\langle \hU_{\bq'\nu'}\hU_{\bq''\nu''} \rangle$ in the right-hand side of Eq.~\eqref{eom_phph}. 
Besides decoherence, this quantity
incorporates the nonlinear phononic driving force \cite{Juraschek2018,JuraschekFechner2017,Mankowsky2017} as well as fluctuation of atomic vibrations (incoherent phonon) \cite{Caruso2023}. 
The dynamics of $\langle \hU_{\bq'\nu'}\hU_{\bq''\nu''} \rangle$ can be determined by generalizing the  
Heisenberg equation of motion from Eq.~\eqref{eq:HEOM}:

\begin{align}\label{eq:bbgkry3}
    &\frac{\partial^2  \langle \hU_{\bq'\nu'}\hU_{\bq''\nu''} \rangle}{\partial t^2} +(\omega_{\bq'\nu'}^2+\omega_{\bq''\nu''}^2) \langle \hU_{\bq'\nu'}\hU_{\bq''\nu''} \rangle = \nonumber \\
    &2\langle \hP_{\bq'\nu'}\hP_{\bq''\nu''} \rangle - \frac{1}{2}\sum_{1,2}V_{-\bq'\nu'\bq_1\nu_1\bq_2\nu_2}^{(3)} \langle\hU_{\bq_1\nu_1}\hU_{\bq_2\nu_2}\hU_{\bq''\nu''} \rangle 
    \nonumber \\ &
    -\frac{1}{2}\sum_{1,2}V_{-\bq''\nu''\bq_1\nu_1\bq_2\nu_2}^{(3)} \langle\hU_{\bq_1\nu_1}\hU_{\bq_2\nu_2}\hU_{\bq'\nu'} \rangle\quad .
\end{align}
where the right-hand side depends on two-momentum $\langle \hP_{\bq'\nu'}\hP_{\bq''\nu''} \rangle$ and three-phonon correlations. 
Equations~\eqref{eom_phph} and \eqref{eq:bbgkry3} are the first terms of the
Bogoliubov–Born–Green–Kirkwood–Yvon (BBGKY) hierarchy, which recast the
equation of motion for the $n$-th order correlator, in terms of correlators of order $n+1$. 
In order to close the equation, we can derive the equation of motion for
$\langle \hP_{\bq'\nu'}\hP_{\bq''\nu''} \rangle$ and truncate the three-phonon
correlations into the product of two-phonon correlation and the coherent phonon
$\langle \hU_{\bq\nu} \rangle$. To close the equation of motion for $\langle
\hU_{\bq'\nu'}\hU_{\bq''\nu''} \rangle$ and $\langle
\hP_{\bq'\nu'}\hP_{\bq''\nu''} \rangle$ we make use of the momentum conservation ($\bq_1+\bq_2={\bf G}+\bq-\bq''$) and we neglect non-diagonal terms contributing to the 
correlation function. This is equivalent to consider only diagonal self-energy contributions.  
After some algebra, the coherent-phonon equation of motion can be recast 
in a form identical to Eq.~\eqref{eom_Q_eph}, with decoherence rate and frequency renormalization 
given by:
\begin{align}\label{eq:Gammapp}
 \Gamma_{\bq\nu}^{\rm pp} &= -{\rm Im}\Pi_{\bq\nu}^{\rm pp} \quad ,\\
\Omega_{\bq\nu}^2 &= \omega_{\bq\nu}^2+2 \omega_{\bq\nu}{\rm Re}\Pi_{\bq\nu}^{{\rm pp}}\quad. 
\end{align}
$\Pi_{\bq\nu}^{\rm pp}$ is the lowest-order phonon self-energy 
due to the phonon-phonon interaction, defined according to:
\begin{widetext}

\begin{align}\label{phphself}  
\Pi_{\bq\nu}^{\rm pp}  = -\sum_{\bq'\nu',\bq''\nu''} &
\Big|\Psi_{-\bq\nu,\bq'\nu',\bq''\nu''}\Big|^2
  \left[
\frac{(\omega_{\bq'\nu'}+\omega_{\bq''\nu''})(n_{\bq'\nu'}+n_{\bq''\nu''}+1)}{(\omega_{\bq'\nu'}+\omega_{\bq''\nu''})^2-(\omega_{\bq\nu}+i\eta)^2}
\right. + \left.
\frac{(\omega_{\bq'\nu'}-\omega_{\bq''\nu''})(n_{\bq''\nu''}-n_{\bq'\nu'})}{(\omega_{\bq'\nu'}-\omega_{\bq''\nu''})^2-(\omega_{\bq\nu}+i\eta)^2}\right]\quad,
\end{align}
\end{widetext}
where we defined phonon-phonon coupling matrix elements as $\Psi_{\bq\nu,\bq'\nu',\bq''\nu''} = \frac{V^{(3)}_{\bq\nu,\bq'\nu',\bq''\nu''}}{\sqrt{2\omega_{\bq\nu} 2\omega_{\bq'\nu'} 2\omega_{\bq''\nu''}}} $, and $n_{\bq\nu}$ denotes the phonon distribution function.  
Upon replacing $n_{\bq\nu}$ with the equilibrium Bose-Einstein distribution function, 
Eq.~\eqref{phphself} coincides with the phonon self-energy 
obtained within the equilibrium Green function formalization \cite{Lazzeri2003,Cowley1968,Thompson1963,Maradudin_Fein1962,LAX1964487}. 

A detailed derivation of this result is reported in the SM~\cite{SI}.
Overall, these results formally relates the decoherence rate and frequency
renormalization of coherent phonons induced by anharmonic effects to the real
and imaginary parts of the phonon self-energy due phonon-phonon scattering,
respectively, thus, providing the theoretical foundation for the investigation
of phonon decoherence from first principles. 

\begin{figure*}[ht!]
\raggedleft 
\includegraphics[width=1.0\textwidth]{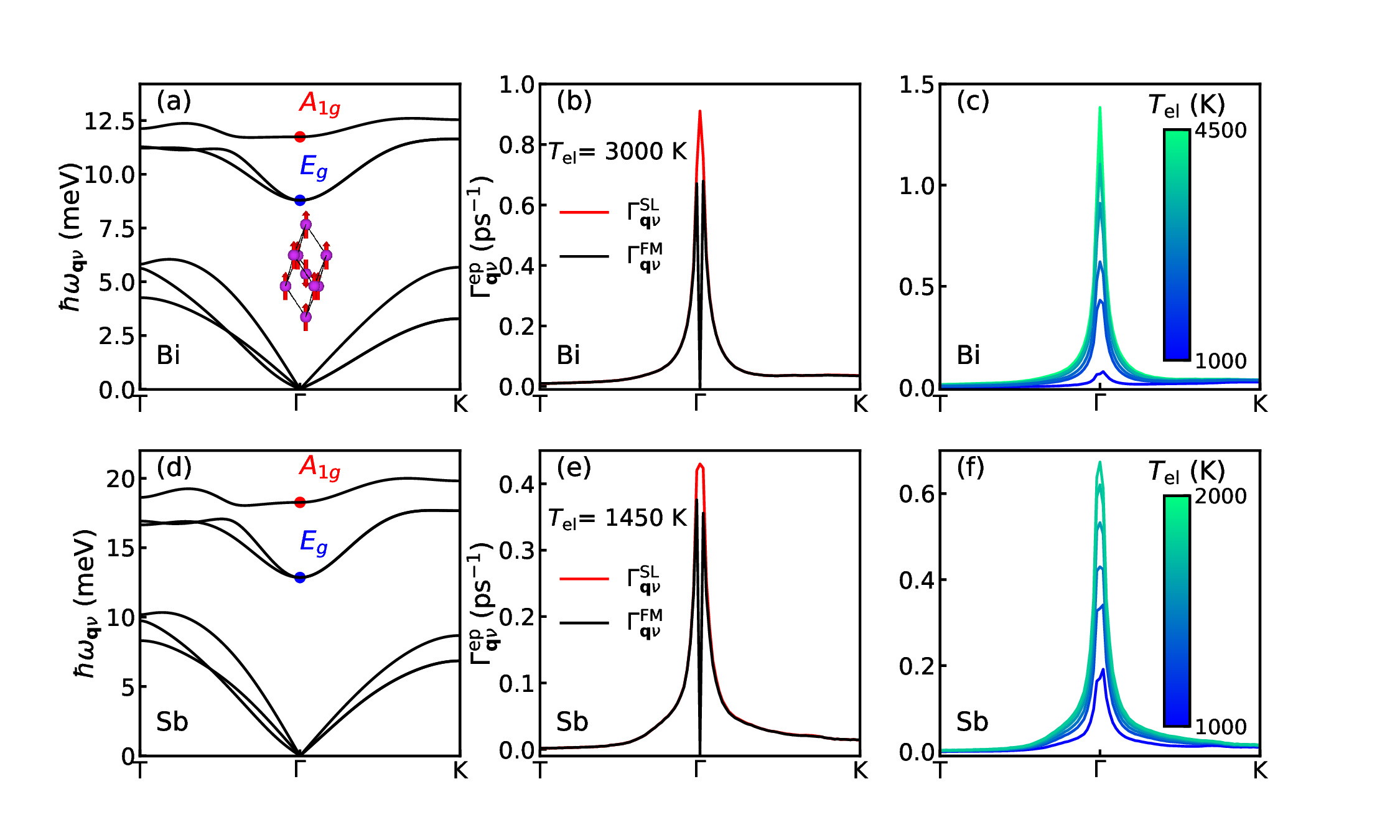}
\caption{ Phonon dispersion of Bi (a) and Sb (d) along the high symmetry path T-$\Gamma$-K of the BZ. The decoherence rate of $A_{1g}$ mode at $\Gamma$ point  along the same high-symmetry path for Bi (b) and Sb (e) with and without the intraband contribution. The electronic temperature is 3000 K for Bi and 1450 K for Sb. The decoherence rate with rising electronic temperature for Bi (c) and Sb (f) which is consistent with the fluence of laser pulse employed in the experiments.}
\label{fig:2}
\end{figure*}

\section{Phonon decoherence in elemental semimetals}

In elemental semimetals, such as  bismuth (Bi) and antimony (Sb), coherent
phonons have been comprehensively characterized via pump-probe optical and
photoemission experiments 
\cite{Hase1998,Hase2015,Teitelbaum2018,Cheng2018,Ishioka2024,EmeisJauernik2024}. 
These works revealed a strong
dependence of decoherence on temperature \cite{Hase1998,Hase2015}
and pump fluence \cite{Teitelbaum2018,EmeisJauernik2024}, making Sb and Bi an ideal platform
to assess the accuracy of the first-principles approach 
developed in Sec.~\ref{sec:theo}. 
In the following, we thus proceed to investigate the decoherence rate of 
coherent phonons in Bi and Sb based on ab initio simulations of 
the electron-phonon and phonon-phonon interactions. 
 All calculations are conducted with the {\tt EPW}
code \cite{bib:epw,Giustino2007} which is part of {\tt Quantum
Espresso} \cite{Giannozzi2017} and is interfaced with {\tt Wannier90}
\cite{pizzi2020wannier90}.
Computational details are reported  in the SM \cite{SI}.

In their low-temperature phase, Bi and Sb crystallize in a rhombohedral
structure belonging to space group $R\bar{3}m$ (inset of Fig~\ref{fig:2}~(a)), 
which arises from a Peierls
distortion along the trigonal axis of the high-symmetry cubic phase.  The
phonon dispersions of Bi and Sb obtained from density-functional perturbation theory are illustrated in Fig.~\ref{fig:2}~(a) and (d),
respectively. The $A_{1g}$ mode is the dominant coherent vibrational excitation observed in pump-probe experiments \cite{Teitelbaum2018,Ishioka2024,EmeisJauernik2024} 
and it coincides with the
$\Gamma$-point high-energy optical phonon, marked by a red dot in
Figs.~\ref{fig:2}~(a) and (d).  The coherent excitation of the $A_{1g}$ phonon is
induced by the displacive mechanism \cite{Zeiger1992,EmeisJauernik2024,Kuznetsov1994}, which results from
the emergence of a photo-excited carrier density 
coupled to the lattice via electron-phonon coupling. An
ab initio description of the $A_{1g}$ coherent phonon
excitation in Sb via the displacive mechanism 
was recently reported in Ref.~\cite{EmeisJauernik2024}.  While also the $E_{g}$ mode (marked in blue in Figs.~\ref{fig:2}~(a) and (d)) can undergo
coherent excitation, its oscillation
amplitude is much smaller than the $A_{1g}$ mode \cite{Garrett1996,LiChenReis2013}, making an experimental 
assessment of decoherence more challenging. We thus concentrate below on the decoherence rate of the
$A_{1g}$ coherent phonon. 

We begin by discussing the decoherence rate due to the electron-phonon interaction.
In Figs.~\ref{fig:2}~(b) and (e),
we report the decoherence rate $\Gamma_{\bq\nu} ^{\rm FM}$ 
of the high-energy optical phonon of Bi and Sb, respectively, for momenta along the T-$\Gamma$-K high-symmetry path of the Brillouin zone.
$\Gamma_{\bq\nu} ^{\rm FM}$ is 
obtained from Eq.~\eqref{eq:PiNA} by considering the Fan-Migdal approximation (FM) to the phonon self-energy $\Pi_{\bq\nu}^{\rm NA}$ due to electron-phonon coupling. 
To explicitly account for the energy transferred to the electrons  by a pump pulse, we describe electronic occupations via a Fermi-Dirac distribution with an electronic temperature of 3000 K for Bi and 1450 K for Sb. Here and below, the electronic temperatures are chosen to reproduce the experimental conditions of Ref.~\cite{Teitelbaum2018,EmeisJauernik2024}, as discussed in the SM \cite{SI}. 
Figures~\ref{fig:2}~(b) and (e) indicate that the decoherence rate $\Gamma^{\rm FM}_{\bq\nu}$ is peaked around the $\Gamma$ point ($\bq=0$). 
However, as $\Gamma$ is approached, the decoherence rate undergoes a sharp decrease and vanishes exactly at $\Gamma$. 
Since coherent 
phonons are excited precisely at $\Gamma$, one would
expect vanishing linewidths, corresponding to undamped coherent oscillations of the lattice.  

The vanishing of $\Gamma^{\rm FM}_{\bq\nu}$, however, is an artifact of the
Fan-Migdal approximation and it arises from the omission of intraband
transitions, which provide the dominant contribution to the self-energy for $\bq =0$. This result indicates that the  
Fan-Migdal approximation is unsuitable for capturing phonon decoherence. 
This problem is well known in the context of (incoherent) phonon damping
and it has been extensively discussed in the context of Raman linewidths of
MgB$_2 $ \cite{Cappelluti2006,SaittaLazzeri2008,Novko2018, NovkoCaruso2020}.  
A
solution to this issue requires accounting for the contribution of intraband transitions to the phonon damping. This can be achieved either via the inclusion of additional
self-energy diagrams \cite{Novko2018} or via the self-consistent solution of the
Dyson equation for the phonon self-energy \cite{LihmPonce2024}. 
While a self-consistent solution of the electron-phonon problem is not
computationally feasible, recently a self-consistent phonon-linewidth approach
has been proposed for both the electron \cite{LihmPonce2024} and  phonon
self-energies \cite{Park2024}. 
This approach mimics the one-shot solution of the Dyson
equation by accounting exclusively for the change of electron linewidths arising from the imaginary part of the electron self-energy. It successfully  accounts for intraband transitions 
and it lifts the unphysical discontinuity of the phonon  self-energy 
as $\Gamma$ is approached. 

In order to extend self-energy calculations to coherent phonon
decoherence in semimetals, we implemented the  self-consistent linewidth (SL) approximation to the  decoherence rate, denoted as $\Gamma_{\bq\nu}^{\rm SL}$ \cite{Park2024}. As illustrated in the SM, 
it can be expressed as: 
\begin{align}\label{eq:SC}
  \Gamma_{\bq\nu}^{\rm SL} = & \pi\sum_{mn\bk}|g_{mn\nu}(\bk,\bq)|^2 \big[f(\varepsilon_{n\bk})-f(\varepsilon_{n\bk}+\hbar\omega_{\bq\nu})\big]\nonumber  \\
  &
  \quad\quad\quad\quad
  \times\delta_{\gamma}(\enk_{m\bk+\bq}-\enk_{n\bk}-\hbar\omega_{\bq\nu} )\quad,
\end{align}
where  $f(\varepsilon) = \{{\rm exp}[(\varepsilon-\mu)/k_{\rm B}T_{\rm el}]+1\}^{-1}$ denotes the Fermi-Dirac function with chemical potential $\mu$ and $\delta_\gamma$ is a Lorentzian function with  width $\gamma = 15$~meV, representative of the electron linewidths in the vicinity of the Fermi surface. 
In the limit of vanishing electron linewidth ($\gamma \rightarrow 0$), $\delta_\gamma$ 
reduces to the Dirac $\delta$  
and one recovers from Eq.~\eqref{eq:SC}  the decoherence rate in the Fan-Migdal approximation $\Gamma_{\bq\nu}^{\rm FM}$ \cite{Giustino2007,Bonini2007,Lazzeri2006,Park2008} (black line in Figs.~\ref{fig:2}~(b) and (e)).

The decoherence rate $\Gamma_{\bq\nu}^{\rm SL}$ of the high-frequency optical phonon of Bi and Sb obtained from the evaluation of 
Eq.~\eqref{eq:SC} is reported in red in Figs.~\ref{fig:2}~(b) and (e) for momenta along the T-$\Gamma$-K path. 
 Away from $\Gamma$, the decoherence rate $\Gamma_{\bq\nu}^{\rm SL}$  coincides with the result of the Fan-Migdal approximation $\Gamma_{\bq\nu}^{\rm FM}$, however, the two approximations differ significantly at the zone center. 
Specifically, $\Gamma_{\bq\nu}^{\rm SL}$ lifts the unphysical discontinuity at $\Gamma$, leading to finite decoherence rates. 
This is critical for investigating the dissipation of coherent phonons induced by electron-phonon coupling, and  it  enables us to estimate the decoherence rate of the $A_{1g}$ mode of Bi (Sb) to 0.9 ps$^{-1}$ (0.4 ps$^{-1}$) for an effective electronic temperature of $T_{\rm el} = $ 3000 K (1450~K).
To investigate the influence of the electronic excitation on phonon decoherence,   
we report in Figs.~\ref{fig:2}~(c) and (f)  decoherence rate $\Gamma_{\bq\nu}^{\rm SL}$ for electronic temperatures ranging up to {2000} and 4500~K. 
We observe a significant increase of the decoherence rate with increasing electronic temperature. This behavior can be attributed to larger phase space available for phonon-assisted electronic transitions at large electronic temperatures, giving rise to faster decoherence. 

\begin{figure}
\centering
\includegraphics[width=0.5\textwidth]{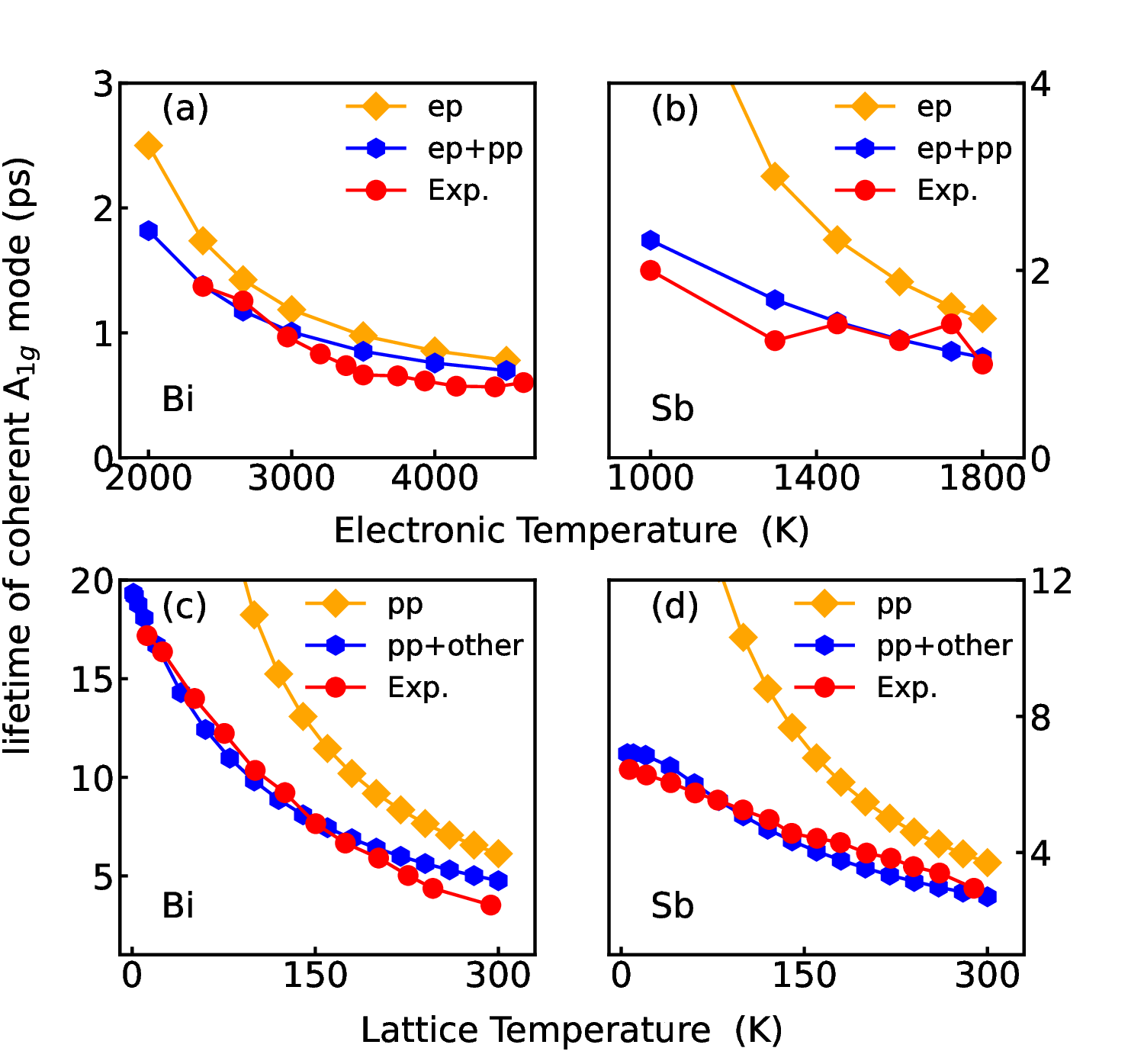}
\caption{Coherent phonon lifetime of the $A_{1g}$ mode for Bi and Sb as a function of electronic temperature [(a)-(b)] and lattice temperature [(c)-(d)]. The experimental data are taken from Ref.~\cite{Teitelbaum2018, Hase1998} for Bi, and Ref.~\cite{EmeisJauernik2024,Hase2015} for Sb.}
\label{fig:3}
\end{figure}

The dependence of the calculated coherent phonon lifetime $\tau_{A_{1g}}^{\rm SL} = (\Gamma^{\rm SL}_{A_{1g}})^{-1}$ for the $A_{1g}$ mode on the 
effective electronic temperature $T_{\rm el}$ is illustrated 
in Figs.~\ref{fig:3}~(a) and (b)
(orange diamonds). 
For comparison, we 
report the experimental lifetime (red circles)
extracted from fluence-dependent pump-probe optical and photoemission experiments of 
Bi ans Sb, respectively 
\cite{Teitelbaum2018,EmeisJauernik2024}. 
The fluence has been expressed in terms of the effective electronic temperature according to the prescription outlined in the SM.  
The dependence of the lifetime {$\tau^{\rm SL}_{A_{1g}}$} on  $T_{\rm el}$ is well approximated
by a linear relation of the form {$\tau^{\rm SL} = \alpha T^{-1}_{\rm el}$}, with 
{$\alpha \sim 2.6$
and $\sim1.6$ 
$\rm ns\cdot K$} for Bi and Sb, respectively. 
Overall, by accounting for the electron-phonon interactions  our simulations capture the {decrease} of the {coherent phonon lifetime} as a function of the electronic temperature (fluence) observed in experiments. 
These findings provide strong evidence that electron-phonon coupling 
is the primary mechanism behind the fluence-dependent decoherence. 
However, calculations based on Eq.~\eqref{eq:SC} alone  systematically overestimate the experimental lifetime. 
Inspection of the decoherence rate (Fig.~S1 in the SM~\cite{SI}) reveals underestimation by a constant offset, suggesting the presence of additional interaction mechanisms -- independent of $T_{\rm el}$ --  contributing to the decoherence rate. As discussed below, this discrepancy can be partially resolved by including the effects of phonon-phonon interactions, which introduce an additional  damping channel beyond electron-phonon coupling.

The contribution of phonon-phonon scattering to the decoherence rate, $\Gamma_{\bq\nu}^{\rm pp}$, can be accounted for through  Eqs.~\eqref{eq:Gammapp} and \eqref{phphself}, leading to:
\begin{align} \label{phlw_pp}
    \Gamma_{\bq\nu}^{\rm pp}  = & \frac{\pi}{2}\sum_{\bq'\bq''\nu'\nu''}|\Psi_{-\bq,\bq',\bq''}^{\nu,\nu',\nu''}|^2  \\ 
    \times&\Big[ (n_{\bq'\nu'}+n_{\bq''\nu''}+1)\delta(\omega_{\bq'\nu'}+\omega_{\bq''\nu''}-\omega_{\bq\nu}) \nonumber  \\
    &+ 2(n_{\bq''\nu''}-n_{\bq'\nu'}) \delta(\omega_{\bq'\nu'}-\omega_{\bq''\nu''}-\omega_{\bq\nu})\Big]\quad.\nonumber
\end{align}
Numerical evaluation of $\Gamma^{\rm pp}_{A_{1g}}$ at room temperature ($T_{\rm ph } =300$~K) for the $A_{1g}$ mode of Sb and Bi yields 0.26 and 0.15~ps$^{-1}$, respectively. 
According to Matthiessen's rule, the total {lifetime $\tau^{\rm ep+pp}_{A_{1g}}$} can be obtained by adding the contributions due to electron-phonon and phonon-phonon scattering: 
$\tau^{\rm ep+pp}_{A_{1g}} = \left[\Gamma^{\rm SL}_{A_{1g}}(T_{\rm el}) + \Gamma^{\rm pp}_{A_{1g}}(T_{\rm ph}) \right]^{-1}$.
Since  $\Gamma^{\rm pp}_{A_{1g}}$ does not depend on the electronic occupations $f_{n\bk}$, we neglect its dependence on $T_{\rm el}$. 
This approximation omits possible changes of the phonon distribution $n_{\bq\nu}$ induced by electronic excitations \cite{Caruso2021,TDBE2,PanCaruso2023,PanCaruso2024}, and their effects of the phonon-phonon decoherence rate. 
The total lifetime $\tau^{\rm ep+pp}_{A_{1g}}$, 
marked  by blue hexagons in 
Fig.~\ref{fig:3}~(a) and (b),  
reduces the discrepancy between experiments and theory, 
indicating that both electron-phonon and phonon scattering 
can contribute to decoherence. 

To illustrate the influence of phonon-phonon interaction on 
decoherence,
we report in Figs.~\ref{fig:3}~(c) and (d) the 
quantity $ \tau_{A_{1g}}^{\rm pp}$
for lattice temperatures $T_{\rm ph}$ ranging between 0 and 300~K (orange diamonds).
The temperature-dependent {coherent phonon lifetime} extracted from pump-probe experiments at constant fluence are shown as red circles \cite{Hase1998,Hase2015}. 
Our calculations accurately reproduce the {decrease of lifetime} with lattice temperature, in agreement with experiments. The temperature dependence is well captured by the fitting function {$\tau^{\rm pp}_{A_{1g}} \simeq \tau_{\rm 0}^{\rm pp}[2n(T_{\rm ph})+1]^{-1}$}, where $n(T_{\rm ph})$ is the Bose-Einstein distribution, which accounts for the increasing phonon population with temperature. After an initial nonlinear transient at low temperatures, $n(T_{\rm ph})$ varies linearly with $T_{\rm ph}$ and the lifetime is well described by the fitting function  
{$\tau^{\rm pp}_{A_{1g}} = [\Gamma_0 +\alpha_{\rm anh} T_{\rm ph}]^{-1}$}, 
with $\alpha_{\rm anh} = 5.0 \times 10^{-4}$ for Bi and $\alpha_{\rm anh} = 7.6\times 10^{-4}$ for Sb. 

Simulations of 
$\Gamma^{\rm pp}_{A_{1g}}$ underestimate 
the experiments by a constant  
offset, as show in Fig.~S1 in the SM~\cite{SI}, indicating the presence of additional contributions 
to the decoherence rates that do not depend significantly 
on the lattice temperature. These effects were attributed to 
electron-phonon  and impurity scattering, 
and they were estimated to be 0.05 ps$^{-1}$ \cite{Hase1998} for Bi and 0.1 ps$^{-1}$ for Sb \cite{Hase2015}.  
To account for these contributions, we introduce an effective constant decoherence rate $\Gamma^{\rm eff}_{A_{1g}}$, leading to a 
total decoherence rate $\Gamma^{\rm tot}_{A_{1g}} =
\Gamma^{\rm pp}_{A_{1g}} + \Gamma^{\rm eff}_{A_{1g}}$.
The total {lifetime} $\tau^{\rm tot}_{A_{1g}} $ is reported in Fig.~\ref{fig:3}~(c)-(d) (blue hexagons), and its 
improved agreement with experiments provides a 
robust interpretation of the residual discrepancy. 
Overall, these findings provide strong evidence that the 
temperature dependence of decoherence is governed 
by phonon-phonon scattering.

\section{Conclusions}
In conclusion, we investigated the process of phonon decoherence in crystalline
solids on the basis of a first-principles many-body framework.  Starting from a
many-body formulation of the lattice equation of motion, we obtain
explicit expressions for the decoherence rate and frequency renormalization of
coherent phonons arising from electron-phonon and phonon-phonon coupling. These
results rigorously link the process of phonon decoherence to the phonon
self-energy. We validate this approach by conducting first-principles
calculations of phonon decoherence induced by electron-phonon and
phonon-phonon interactions.  Specifically, we estimated the fluence and
temperature dependence of the phonon decoherence time for the elemental
semimetals Bi and Sb. The good quantitative agreement with recent pump-probe
experimental data corroborate the theoretical approach introduced here.
These findings further indicate that far from equilibrium both
electron-phonon and phonon-phonon scattering can  
contribute substantially to the process of phonon decoherence. 
Besides advancing the fundamental understanding of the non-equilibrium 
lattice dynamics, these results constitute the basis to understand and assess the 
characteristic timescales for phonon decoherence in driven solids, with 
implications for light-induced structural control and phase transitions.  

\section*{acknowledgement}
This work is funded by the Deutsche Forschungsgemeinschaft (DFG), Projects No. 443988403 and No. 499426961. The authors gratefully acknowledge the
computing time provided by the high-performance computer Lichtenberg at the NHR Centers NHR4CES at TU
Darmstadt (Project p0021280). 
We are grateful for the fruitful discussions with Elia Stocco, Cheol-Hwan Park and Andrea Marini.


\begin{thebibliography}{72}%
\makeatletter
\providecommand \@ifxundefined [1]{%
 \@ifx{#1\undefined}
}%
\providecommand \@ifnum [1]{%
 \ifnum #1\expandafter \@firstoftwo
 \else \expandafter \@secondoftwo
 \fi
}%
\providecommand \@ifx [1]{%
 \ifx #1\expandafter \@firstoftwo
 \else \expandafter \@secondoftwo
 \fi
}%
\providecommand \natexlab [1]{#1}%
\providecommand \enquote  [1]{``#1''}%
\providecommand \bibnamefont  [1]{#1}%
\providecommand \bibfnamefont [1]{#1}%
\providecommand \citenamefont [1]{#1}%
\providecommand \href@noop [0]{\@secondoftwo}%
\providecommand \href [0]{\begingroup \@sanitize@url \@href}%
\providecommand \@href[1]{\@@startlink{#1}\@@href}%
\providecommand \@@href[1]{\endgroup#1\@@endlink}%
\providecommand \@sanitize@url [0]{\catcode `\\12\catcode `\$12\catcode `\&12\catcode `\#12\catcode `\^12\catcode `\_12\catcode `\%12\relax}%
\providecommand \@@startlink[1]{}%
\providecommand \@@endlink[0]{}%
\providecommand \url  [0]{\begingroup\@sanitize@url \@url }%
\providecommand \@url [1]{\endgroup\@href {#1}{\urlprefix }}%
\providecommand \urlprefix  [0]{URL }%
\providecommand \Eprint [0]{\href }%
\providecommand \doibase [0]{https://doi.org/}%
\providecommand \selectlanguage [0]{\@gobble}%
\providecommand \bibinfo  [0]{\@secondoftwo}%
\providecommand \bibfield  [0]{\@secondoftwo}%
\providecommand \translation [1]{[#1]}%
\providecommand \BibitemOpen [0]{}%
\providecommand \bibitemStop [0]{}%
\providecommand \bibitemNoStop [0]{.\EOS\space}%
\providecommand \EOS [0]{\spacefactor3000\relax}%
\providecommand \BibitemShut  [1]{\csname bibitem#1\endcsname}%
\let\auto@bib@innerbib\@empty
\bibitem [{\citenamefont {Horstmann}\ \emph {et~al.}(2020)\citenamefont {Horstmann}, \citenamefont {B{\"o}ckmann}, \citenamefont {Wit}, \citenamefont {Kurtz}, \citenamefont {Storeck},\ and\ \citenamefont {Ropers}}]{horstmann2020coherent}%
  \BibitemOpen
  \bibfield  {author} {\bibinfo {author} {\bibfnamefont {J.~G.}\ \bibnamefont {Horstmann}}, \bibinfo {author} {\bibfnamefont {H.}~\bibnamefont {B{\"o}ckmann}}, \bibinfo {author} {\bibfnamefont {B.}~\bibnamefont {Wit}}, \bibinfo {author} {\bibfnamefont {F.}~\bibnamefont {Kurtz}}, \bibinfo {author} {\bibfnamefont {G.}~\bibnamefont {Storeck}},\ and\ \bibinfo {author} {\bibfnamefont {C.}~\bibnamefont {Ropers}},\ }\bibfield  {title} {\bibinfo {title} {Coherent control of a surface structural phase transition},\ }\href {https://doi.org/10.1038/s41586-020-2440-4} {\bibfield  {journal} {\bibinfo  {journal} {Nature}\ }\textbf {\bibinfo {volume} {583}},\ \bibinfo {pages} {232} (\bibinfo {year} {2020})}\BibitemShut {NoStop}%
\bibitem [{\citenamefont {Zhang}\ \emph {et~al.}(2019)\citenamefont {Zhang}, \citenamefont {Wang}, \citenamefont {Li}, \citenamefont {Shi}, \citenamefont {Wu}, \citenamefont {Lin}, \citenamefont {Zhang}, \citenamefont {Liu}, \citenamefont {Liu}, \citenamefont {Wang}, \citenamefont {Dong},\ and\ \citenamefont {Wang}}]{ZhangWang2019}%
  \BibitemOpen
  \bibfield  {author} {\bibinfo {author} {\bibfnamefont {M.~Y.}\ \bibnamefont {Zhang}}, \bibinfo {author} {\bibfnamefont {Z.~X.}\ \bibnamefont {Wang}}, \bibinfo {author} {\bibfnamefont {Y.~N.}\ \bibnamefont {Li}}, \bibinfo {author} {\bibfnamefont {L.~Y.}\ \bibnamefont {Shi}}, \bibinfo {author} {\bibfnamefont {D.}~\bibnamefont {Wu}}, \bibinfo {author} {\bibfnamefont {T.}~\bibnamefont {Lin}}, \bibinfo {author} {\bibfnamefont {S.~J.}\ \bibnamefont {Zhang}}, \bibinfo {author} {\bibfnamefont {Y.~Q.}\ \bibnamefont {Liu}}, \bibinfo {author} {\bibfnamefont {Q.~M.}\ \bibnamefont {Liu}}, \bibinfo {author} {\bibfnamefont {J.}~\bibnamefont {Wang}}, \bibinfo {author} {\bibfnamefont {T.}~\bibnamefont {Dong}},\ and\ \bibinfo {author} {\bibfnamefont {N.~L.}\ \bibnamefont {Wang}},\ }\bibfield  {title} {\bibinfo {title} {Light-induced subpicosecond lattice symmetry switch in {MoTe$_2$}},\ }\href {https://doi.org/10.1103/PhysRevX.9.021036} {\bibfield  {journal} {\bibinfo  {journal} {Phys. Rev. X}\ }\textbf {\bibinfo {volume}
  {9}},\ \bibinfo {pages} {021036} (\bibinfo {year} {2019})}\BibitemShut {NoStop}%
\bibitem [{\citenamefont {Qi}\ \emph {et~al.}(2022)\citenamefont {Qi}, \citenamefont {Guan}, \citenamefont {Zahn}, \citenamefont {Vasileiadis}, \citenamefont {Seiler}, \citenamefont {Windsor}, \citenamefont {Zhao}, \citenamefont {Meng},\ and\ \citenamefont {Ernstorfer}}]{QiGuanZahn2022}%
  \BibitemOpen
  \bibfield  {author} {\bibinfo {author} {\bibfnamefont {Y.}~\bibnamefont {Qi}}, \bibinfo {author} {\bibfnamefont {M.}~\bibnamefont {Guan}}, \bibinfo {author} {\bibfnamefont {D.}~\bibnamefont {Zahn}}, \bibinfo {author} {\bibfnamefont {T.}~\bibnamefont {Vasileiadis}}, \bibinfo {author} {\bibfnamefont {H.}~\bibnamefont {Seiler}}, \bibinfo {author} {\bibfnamefont {Y.~W.}\ \bibnamefont {Windsor}}, \bibinfo {author} {\bibfnamefont {H.}~\bibnamefont {Zhao}}, \bibinfo {author} {\bibfnamefont {S.}~\bibnamefont {Meng}},\ and\ \bibinfo {author} {\bibfnamefont {R.}~\bibnamefont {Ernstorfer}},\ }\bibfield  {title} {\bibinfo {title} {Traversing double-well potential energy surfaces: Photoinduced concurrent intralayer and interlayer structural transitions in {XTe$_2$} ({X = Mo, W})},\ }\href {https://doi.org/10.1021/acsnano.2c03809} {\bibfield  {journal} {\bibinfo  {journal} {ACS Nano}\ }\textbf {\bibinfo {volume} {16}},\ \bibinfo {pages} {11124} (\bibinfo {year} {2022})}\BibitemShut {NoStop}%
\bibitem [{\citenamefont {Guan}\ \emph {et~al.}(2022)\citenamefont {Guan}, \citenamefont {Liu}, \citenamefont {Chen}, \citenamefont {Li}, \citenamefont {Qi}, \citenamefont {Yang}, \citenamefont {You},\ and\ \citenamefont {Meng}}]{GuanLiuChen2022}%
  \BibitemOpen
  \bibfield  {author} {\bibinfo {author} {\bibfnamefont {M.-X.}\ \bibnamefont {Guan}}, \bibinfo {author} {\bibfnamefont {X.-B.}\ \bibnamefont {Liu}}, \bibinfo {author} {\bibfnamefont {D.-Q.}\ \bibnamefont {Chen}}, \bibinfo {author} {\bibfnamefont {X.-Y.}\ \bibnamefont {Li}}, \bibinfo {author} {\bibfnamefont {Y.-P.}\ \bibnamefont {Qi}}, \bibinfo {author} {\bibfnamefont {Q.}~\bibnamefont {Yang}}, \bibinfo {author} {\bibfnamefont {P.-W.}\ \bibnamefont {You}},\ and\ \bibinfo {author} {\bibfnamefont {S.}~\bibnamefont {Meng}},\ }\bibfield  {title} {\bibinfo {title} {Optical control of multistage phase transition via phonon coupling in {MoTe$_2$}},\ }\href {https://doi.org/10.1103/PhysRevLett.128.015702} {\bibfield  {journal} {\bibinfo  {journal} {Phys. Rev. Lett.}\ }\textbf {\bibinfo {volume} {128}},\ \bibinfo {pages} {015702} (\bibinfo {year} {2022})}\BibitemShut {NoStop}%
\bibitem [{\citenamefont {Li}\ \emph {et~al.}(2019)\citenamefont {Li}, \citenamefont {Qiu}, \citenamefont {Zhang}, \citenamefont {Baldini}, \citenamefont {Lu}, \citenamefont {Rappe},\ and\ \citenamefont {Nelson}}]{LiQiu2019}%
  \BibitemOpen
  \bibfield  {author} {\bibinfo {author} {\bibfnamefont {X.}~\bibnamefont {Li}}, \bibinfo {author} {\bibfnamefont {T.}~\bibnamefont {Qiu}}, \bibinfo {author} {\bibfnamefont {J.}~\bibnamefont {Zhang}}, \bibinfo {author} {\bibfnamefont {E.}~\bibnamefont {Baldini}}, \bibinfo {author} {\bibfnamefont {J.}~\bibnamefont {Lu}}, \bibinfo {author} {\bibfnamefont {A.~M.}\ \bibnamefont {Rappe}},\ and\ \bibinfo {author} {\bibfnamefont {K.~A.}\ \bibnamefont {Nelson}},\ }\bibfield  {title} {\bibinfo {title} {Terahertz field induced ferroelectricity in quantum paraelectric {SrTiO$_3$}},\ }\href {https://doi.org/10.1126/science.aaw4913} {\bibfield  {journal} {\bibinfo  {journal} {Science}\ }\textbf {\bibinfo {volume} {364}},\ \bibinfo {pages} {1079} (\bibinfo {year} {2019})}\BibitemShut {NoStop}%
\bibitem [{\citenamefont {de~la Torre}\ \emph {et~al.}(2021)\citenamefont {de~la Torre}, \citenamefont {Kennes}, \citenamefont {Claassen}, \citenamefont {Gerber}, \citenamefont {McIver},\ and\ \citenamefont {Sentef}}]{delaTorre2021}%
  \BibitemOpen
  \bibfield  {author} {\bibinfo {author} {\bibfnamefont {A.}~\bibnamefont {de~la Torre}}, \bibinfo {author} {\bibfnamefont {D.~M.}\ \bibnamefont {Kennes}}, \bibinfo {author} {\bibfnamefont {M.}~\bibnamefont {Claassen}}, \bibinfo {author} {\bibfnamefont {S.}~\bibnamefont {Gerber}}, \bibinfo {author} {\bibfnamefont {J.~W.}\ \bibnamefont {McIver}},\ and\ \bibinfo {author} {\bibfnamefont {M.~A.}\ \bibnamefont {Sentef}},\ }\bibfield  {title} {\bibinfo {title} {Colloquium: Nonthermal pathways to ultrafast control in quantum materials},\ }\href {https://doi.org/10.1103/RevModPhys.93.041002} {\bibfield  {journal} {\bibinfo  {journal} {Rev. Mod. Phys.}\ }\textbf {\bibinfo {volume} {93}},\ \bibinfo {pages} {041002} (\bibinfo {year} {2021})}\BibitemShut {NoStop}%
\bibitem [{\citenamefont {F{\"o}rst}\ \emph {et~al.}(2011)\citenamefont {F{\"o}rst}, \citenamefont {Manzoni}, \citenamefont {Kaiser}, \citenamefont {Tomioka}, \citenamefont {Tokura}, \citenamefont {Merlin},\ and\ \citenamefont {Cavalleri}}]{Foerst2011}%
  \BibitemOpen
  \bibfield  {author} {\bibinfo {author} {\bibfnamefont {M.}~\bibnamefont {F{\"o}rst}}, \bibinfo {author} {\bibfnamefont {C.}~\bibnamefont {Manzoni}}, \bibinfo {author} {\bibfnamefont {S.}~\bibnamefont {Kaiser}}, \bibinfo {author} {\bibfnamefont {Y.}~\bibnamefont {Tomioka}}, \bibinfo {author} {\bibfnamefont {Y.}~\bibnamefont {Tokura}}, \bibinfo {author} {\bibfnamefont {R.}~\bibnamefont {Merlin}},\ and\ \bibinfo {author} {\bibfnamefont {A.}~\bibnamefont {Cavalleri}},\ }\bibfield  {title} {\bibinfo {title} {Nonlinear phononics as an ultrafast route to lattice control},\ }\href {https://doi.org/10.1038/nphys2055} {\bibfield  {journal} {\bibinfo  {journal} {Nat. Phys.}\ }\textbf {\bibinfo {volume} {7}},\ \bibinfo {pages} {854} (\bibinfo {year} {2011})}\BibitemShut {NoStop}%
\bibitem [{\citenamefont {Guan}\ \emph {et~al.}(2023)\citenamefont {Guan}, \citenamefont {Chen}, \citenamefont {Chen}, \citenamefont {Yao},\ and\ \citenamefont {Meng}}]{GuanChen2023}%
  \BibitemOpen
  \bibfield  {author} {\bibinfo {author} {\bibfnamefont {M.}~\bibnamefont {Guan}}, \bibinfo {author} {\bibfnamefont {D.}~\bibnamefont {Chen}}, \bibinfo {author} {\bibfnamefont {Q.}~\bibnamefont {Chen}}, \bibinfo {author} {\bibfnamefont {Y.}~\bibnamefont {Yao}},\ and\ \bibinfo {author} {\bibfnamefont {S.}~\bibnamefont {Meng}},\ }\bibfield  {title} {\bibinfo {title} {Coherent phonon assisted ultrafast order-parameter reversal and hidden metallic state in {Ta$_2$NiSe$_5$}},\ }\href {https://doi.org/10.1103/PhysRevLett.131.256503} {\bibfield  {journal} {\bibinfo  {journal} {Phys. Rev. Lett.}\ }\textbf {\bibinfo {volume} {131}},\ \bibinfo {pages} {256503} (\bibinfo {year} {2023})}\BibitemShut {NoStop}%
\bibitem [{\citenamefont {Juraschek}\ \emph {et~al.}(2017)\citenamefont {Juraschek}, \citenamefont {Fechner},\ and\ \citenamefont {Spaldin}}]{JuraschekFechner2017}%
  \BibitemOpen
  \bibfield  {author} {\bibinfo {author} {\bibfnamefont {D.~M.}\ \bibnamefont {Juraschek}}, \bibinfo {author} {\bibfnamefont {M.}~\bibnamefont {Fechner}},\ and\ \bibinfo {author} {\bibfnamefont {N.~A.}\ \bibnamefont {Spaldin}},\ }\bibfield  {title} {\bibinfo {title} {Ultrafast structure switching through nonlinear phononics},\ }\href {https://doi.org/10.1103/PhysRevLett.118.054101} {\bibfield  {journal} {\bibinfo  {journal} {Phys. Rev. Lett.}\ }\textbf {\bibinfo {volume} {118}},\ \bibinfo {pages} {054101} (\bibinfo {year} {2017})}\BibitemShut {NoStop}%
\bibitem [{\citenamefont {Juraschek}\ \emph {et~al.}(2020)\citenamefont {Juraschek}, \citenamefont {Meier},\ and\ \citenamefont {Narang}}]{JuraschekMeier2020}%
  \BibitemOpen
  \bibfield  {author} {\bibinfo {author} {\bibfnamefont {D.~M.}\ \bibnamefont {Juraschek}}, \bibinfo {author} {\bibfnamefont {Q.~N.}\ \bibnamefont {Meier}},\ and\ \bibinfo {author} {\bibfnamefont {P.}~\bibnamefont {Narang}},\ }\bibfield  {title} {\bibinfo {title} {Parametric excitation of an optically silent goldstone-like phonon mode},\ }\href {https://doi.org/10.1103/PhysRevLett.124.117401} {\bibfield  {journal} {\bibinfo  {journal} {Phys. Rev. Lett.}\ }\textbf {\bibinfo {volume} {124}},\ \bibinfo {pages} {117401} (\bibinfo {year} {2020})}\BibitemShut {NoStop}%
\bibitem [{\citenamefont {Scholz}\ \emph {et~al.}(1993)\citenamefont {Scholz}, \citenamefont {Pfeifer},\ and\ \citenamefont {Kurz}}]{Scholz1993}%
  \BibitemOpen
  \bibfield  {author} {\bibinfo {author} {\bibfnamefont {R.}~\bibnamefont {Scholz}}, \bibinfo {author} {\bibfnamefont {T.}~\bibnamefont {Pfeifer}},\ and\ \bibinfo {author} {\bibfnamefont {H.}~\bibnamefont {Kurz}},\ }\bibfield  {title} {\bibinfo {title} {Density-matrix theory of coherent phonon oscillations in germanium},\ }\href {https://doi.org/10.1103/PhysRevB.47.16229} {\bibfield  {journal} {\bibinfo  {journal} {Phys. Rev. B}\ }\textbf {\bibinfo {volume} {47}},\ \bibinfo {pages} {16229} (\bibinfo {year} {1993})}\BibitemShut {NoStop}%
\bibitem [{\citenamefont {Garrett}\ \emph {et~al.}(1996)\citenamefont {Garrett}, \citenamefont {Albrecht}, \citenamefont {Whitaker},\ and\ \citenamefont {Merlin}}]{Garrett1996}%
  \BibitemOpen
  \bibfield  {author} {\bibinfo {author} {\bibfnamefont {G.~A.}\ \bibnamefont {Garrett}}, \bibinfo {author} {\bibfnamefont {T.~F.}\ \bibnamefont {Albrecht}}, \bibinfo {author} {\bibfnamefont {J.~F.}\ \bibnamefont {Whitaker}},\ and\ \bibinfo {author} {\bibfnamefont {R.}~\bibnamefont {Merlin}},\ }\bibfield  {title} {\bibinfo {title} {Coherent {THz} phonons driven by light pulses and the {Sb} problem: What is the mechanism?},\ }\href {https://doi.org/10.1103/PhysRevLett.77.3661} {\bibfield  {journal} {\bibinfo  {journal} {Phys. Rev. Lett.}\ }\textbf {\bibinfo {volume} {77}},\ \bibinfo {pages} {3661} (\bibinfo {year} {1996})}\BibitemShut {NoStop}%
\bibitem [{\citenamefont {Qi}\ \emph {et~al.}(2009)\citenamefont {Qi}, \citenamefont {Shin}, \citenamefont {Yeh}, \citenamefont {Nelson},\ and\ \citenamefont {Rappe}}]{QiShin2009}%
  \BibitemOpen
  \bibfield  {author} {\bibinfo {author} {\bibfnamefont {T.}~\bibnamefont {Qi}}, \bibinfo {author} {\bibfnamefont {Y.-H.}\ \bibnamefont {Shin}}, \bibinfo {author} {\bibfnamefont {K.-L.}\ \bibnamefont {Yeh}}, \bibinfo {author} {\bibfnamefont {K.~A.}\ \bibnamefont {Nelson}},\ and\ \bibinfo {author} {\bibfnamefont {A.~M.}\ \bibnamefont {Rappe}},\ }\bibfield  {title} {\bibinfo {title} {Collective coherent control: Synchronization of polarization in ferroelectric {PbTiO}$_{3}$ by shaped {THz} fields},\ }\href {https://doi.org/10.1103/PhysRevLett.102.247603} {\bibfield  {journal} {\bibinfo  {journal} {Phys. Rev. Lett.}\ }\textbf {\bibinfo {volume} {102}},\ \bibinfo {pages} {247603} (\bibinfo {year} {2009})}\BibitemShut {NoStop}%
\bibitem [{\citenamefont {Merlin}(1997)}]{MERLIN1997207}%
  \BibitemOpen
  \bibfield  {author} {\bibinfo {author} {\bibfnamefont {R.}~\bibnamefont {Merlin}},\ }\bibfield  {title} {\bibinfo {title} {Generating coherent {THz} phonons with light pulses},\ }\href {https://doi.org/https://doi.org/10.1016/S0038-1098(96)00721-1} {\bibfield  {journal} {\bibinfo  {journal} {Solid State Commun.}\ }\textbf {\bibinfo {volume} {102}},\ \bibinfo {pages} {207} (\bibinfo {year} {1997})},\ \bibinfo {note} {highlights in Condensed Matter Physics and Materials Science}\BibitemShut {NoStop}%
\bibitem [{\citenamefont {Juraschek}\ and\ \citenamefont {Maehrlein}(2018)}]{Juraschek2018}%
  \BibitemOpen
  \bibfield  {author} {\bibinfo {author} {\bibfnamefont {D.~M.}\ \bibnamefont {Juraschek}}\ and\ \bibinfo {author} {\bibfnamefont {S.~F.}\ \bibnamefont {Maehrlein}},\ }\bibfield  {title} {\bibinfo {title} {Sum-frequency ionic raman scattering},\ }\href {https://doi.org/10.1103/PhysRevB.97.174302} {\bibfield  {journal} {\bibinfo  {journal} {Phys. Rev. B}\ }\textbf {\bibinfo {volume} {97}},\ \bibinfo {pages} {174302} (\bibinfo {year} {2018})}\BibitemShut {NoStop}%
\bibitem [{\citenamefont {Mankowsky}\ \emph {et~al.}(2017)\citenamefont {Mankowsky}, \citenamefont {von Hoegen}, \citenamefont {F\"orst},\ and\ \citenamefont {Cavalleri}}]{Mankowsky2017}%
  \BibitemOpen
  \bibfield  {author} {\bibinfo {author} {\bibfnamefont {R.}~\bibnamefont {Mankowsky}}, \bibinfo {author} {\bibfnamefont {A.}~\bibnamefont {von Hoegen}}, \bibinfo {author} {\bibfnamefont {M.}~\bibnamefont {F\"orst}},\ and\ \bibinfo {author} {\bibfnamefont {A.}~\bibnamefont {Cavalleri}},\ }\bibfield  {title} {\bibinfo {title} {Ultrafast reversal of the ferroelectric polarization},\ }\href {https://doi.org/10.1103/PhysRevLett.118.197601} {\bibfield  {journal} {\bibinfo  {journal} {Phys. Rev. Lett.}\ }\textbf {\bibinfo {volume} {118}},\ \bibinfo {pages} {197601} (\bibinfo {year} {2017})}\BibitemShut {NoStop}%
\bibitem [{\citenamefont {Teitelbaum}\ \emph {et~al.}(2018)\citenamefont {Teitelbaum}, \citenamefont {Shin}, \citenamefont {Wolfson}, \citenamefont {Cheng}, \citenamefont {Molesky}, \citenamefont {Kandyla},\ and\ \citenamefont {Nelson}}]{Teitelbaum2018}%
  \BibitemOpen
  \bibfield  {author} {\bibinfo {author} {\bibfnamefont {S.~W.}\ \bibnamefont {Teitelbaum}}, \bibinfo {author} {\bibfnamefont {T.}~\bibnamefont {Shin}}, \bibinfo {author} {\bibfnamefont {J.~W.}\ \bibnamefont {Wolfson}}, \bibinfo {author} {\bibfnamefont {Y.-H.}\ \bibnamefont {Cheng}}, \bibinfo {author} {\bibfnamefont {I.~J.~P.}\ \bibnamefont {Molesky}}, \bibinfo {author} {\bibfnamefont {M.}~\bibnamefont {Kandyla}},\ and\ \bibinfo {author} {\bibfnamefont {K.~A.}\ \bibnamefont {Nelson}},\ }\bibfield  {title} {\bibinfo {title} {Real-time observation of a coherent lattice transformation into a high-symmetry phase},\ }\href {https://doi.org/10.1103/PhysRevX.8.031081} {\bibfield  {journal} {\bibinfo  {journal} {Phys. Rev. X}\ }\textbf {\bibinfo {volume} {8}},\ \bibinfo {pages} {031081} (\bibinfo {year} {2018})}\BibitemShut {NoStop}%
\bibitem [{\citenamefont {Caruso}\ and\ \citenamefont {Zacharias}(2023)}]{Caruso2023}%
  \BibitemOpen
  \bibfield  {author} {\bibinfo {author} {\bibfnamefont {F.}~\bibnamefont {Caruso}}\ and\ \bibinfo {author} {\bibfnamefont {M.}~\bibnamefont {Zacharias}},\ }\bibfield  {title} {\bibinfo {title} {Quantum theory of light-driven coherent lattice dynamics},\ }\href {https://doi.org/10.1103/PhysRevB.107.054102} {\bibfield  {journal} {\bibinfo  {journal} {Phys. Rev. B}\ }\textbf {\bibinfo {volume} {107}},\ \bibinfo {pages} {054102} (\bibinfo {year} {2023})}\BibitemShut {NoStop}%
\bibitem [{\citenamefont {Bauer}\ \emph {et~al.}(1998)\citenamefont {Bauer}, \citenamefont {Schmid}, \citenamefont {Pavone},\ and\ \citenamefont {Strauch}}]{BauerSchmid1998}%
  \BibitemOpen
  \bibfield  {author} {\bibinfo {author} {\bibfnamefont {R.}~\bibnamefont {Bauer}}, \bibinfo {author} {\bibfnamefont {A.}~\bibnamefont {Schmid}}, \bibinfo {author} {\bibfnamefont {P.}~\bibnamefont {Pavone}},\ and\ \bibinfo {author} {\bibfnamefont {D.}~\bibnamefont {Strauch}},\ }\bibfield  {title} {\bibinfo {title} {Electron-phonon coupling in the metallic elements al, au, na, and nb: A first-principles study},\ }\href {https://doi.org/10.1103/PhysRevB.57.11276} {\bibfield  {journal} {\bibinfo  {journal} {Phys. Rev. B}\ }\textbf {\bibinfo {volume} {57}},\ \bibinfo {pages} {11276} (\bibinfo {year} {1998})}\BibitemShut {NoStop}%
\bibitem [{\citenamefont {Cappelluti}(2006)}]{Cappelluti2006}%
  \BibitemOpen
  \bibfield  {author} {\bibinfo {author} {\bibfnamefont {E.}~\bibnamefont {Cappelluti}},\ }\bibfield  {title} {\bibinfo {title} {Electron-phonon effects on the raman spectrum in $\mathrm{Mg}{\mathrm{b}}_{2}$},\ }\href {https://doi.org/10.1103/PhysRevB.73.140505} {\bibfield  {journal} {\bibinfo  {journal} {Phys. Rev. B}\ }\textbf {\bibinfo {volume} {73}},\ \bibinfo {pages} {140505} (\bibinfo {year} {2006})}\BibitemShut {NoStop}%
\bibitem [{\citenamefont {Saitta}\ \emph {et~al.}(2008)\citenamefont {Saitta}, \citenamefont {Lazzeri}, \citenamefont {Calandra},\ and\ \citenamefont {Mauri}}]{SaittaLazzeri2008}%
  \BibitemOpen
  \bibfield  {author} {\bibinfo {author} {\bibfnamefont {A.~M.}\ \bibnamefont {Saitta}}, \bibinfo {author} {\bibfnamefont {M.}~\bibnamefont {Lazzeri}}, \bibinfo {author} {\bibfnamefont {M.}~\bibnamefont {Calandra}},\ and\ \bibinfo {author} {\bibfnamefont {F.}~\bibnamefont {Mauri}},\ }\bibfield  {title} {\bibinfo {title} {Giant nonadiabatic effects in layer metals: Raman spectra of intercalated graphite explained},\ }\href {https://doi.org/10.1103/PhysRevLett.100.226401} {\bibfield  {journal} {\bibinfo  {journal} {Phys. Rev. Lett.}\ }\textbf {\bibinfo {volume} {100}},\ \bibinfo {pages} {226401} (\bibinfo {year} {2008})}\BibitemShut {NoStop}%
\bibitem [{\citenamefont {Park}\ \emph {et~al.}(2008)\citenamefont {Park}, \citenamefont {Giustino}, \citenamefont {Cohen},\ and\ \citenamefont {Louie}}]{Park2008}%
  \BibitemOpen
  \bibfield  {author} {\bibinfo {author} {\bibfnamefont {C.-H.}\ \bibnamefont {Park}}, \bibinfo {author} {\bibfnamefont {F.}~\bibnamefont {Giustino}}, \bibinfo {author} {\bibfnamefont {M.~L.}\ \bibnamefont {Cohen}},\ and\ \bibinfo {author} {\bibfnamefont {S.~G.}\ \bibnamefont {Louie}},\ }\bibfield  {title} {\bibinfo {title} {Electron-phonon interactions in graphene, bilayer graphene, and graphite},\ }\href {https://doi.org/10.1021/nl801884n} {\bibfield  {journal} {\bibinfo  {journal} {Nano Lett.}\ }\textbf {\bibinfo {volume} {8}},\ \bibinfo {pages} {4229} (\bibinfo {year} {2008})}\BibitemShut {NoStop}%
\bibitem [{\citenamefont {Lazzeri}\ \emph {et~al.}(2006)\citenamefont {Lazzeri}, \citenamefont {Piscanec}, \citenamefont {Mauri}, \citenamefont {Ferrari},\ and\ \citenamefont {Robertson}}]{Lazzeri2006}%
  \BibitemOpen
  \bibfield  {author} {\bibinfo {author} {\bibfnamefont {M.}~\bibnamefont {Lazzeri}}, \bibinfo {author} {\bibfnamefont {S.}~\bibnamefont {Piscanec}}, \bibinfo {author} {\bibfnamefont {F.}~\bibnamefont {Mauri}}, \bibinfo {author} {\bibfnamefont {A.~C.}\ \bibnamefont {Ferrari}},\ and\ \bibinfo {author} {\bibfnamefont {J.}~\bibnamefont {Robertson}},\ }\bibfield  {title} {\bibinfo {title} {Phonon linewidths and electron-phonon coupling in graphite and nanotubes},\ }\href {https://doi.org/10.1103/PhysRevB.73.155426} {\bibfield  {journal} {\bibinfo  {journal} {Phys. Rev. B}\ }\textbf {\bibinfo {volume} {73}},\ \bibinfo {pages} {155426} (\bibinfo {year} {2006})}\BibitemShut {NoStop}%
\bibitem [{\citenamefont {Calandra}\ \emph {et~al.}(2010)\citenamefont {Calandra}, \citenamefont {Profeta},\ and\ \citenamefont {Mauri}}]{Calandra2010}%
  \BibitemOpen
  \bibfield  {author} {\bibinfo {author} {\bibfnamefont {M.}~\bibnamefont {Calandra}}, \bibinfo {author} {\bibfnamefont {G.}~\bibnamefont {Profeta}},\ and\ \bibinfo {author} {\bibfnamefont {F.}~\bibnamefont {Mauri}},\ }\bibfield  {title} {\bibinfo {title} {Adiabatic and nonadiabatic phonon dispersion in a wannier function approach},\ }\href {https://doi.org/10.1103/PhysRevB.82.165111} {\bibfield  {journal} {\bibinfo  {journal} {Phys. Rev. B}\ }\textbf {\bibinfo {volume} {82}},\ \bibinfo {pages} {165111} (\bibinfo {year} {2010})}\BibitemShut {NoStop}%
\bibitem [{\citenamefont {Giustino}\ \emph {et~al.}(2007)\citenamefont {Giustino}, \citenamefont {Cohen},\ and\ \citenamefont {Louie}}]{Giustino2007}%
  \BibitemOpen
  \bibfield  {author} {\bibinfo {author} {\bibfnamefont {F.}~\bibnamefont {Giustino}}, \bibinfo {author} {\bibfnamefont {M.~L.}\ \bibnamefont {Cohen}},\ and\ \bibinfo {author} {\bibfnamefont {S.~G.}\ \bibnamefont {Louie}},\ }\bibfield  {title} {\bibinfo {title} {Electron-phonon interaction using wannier functions},\ }\href {https://doi.org/10.1103/PhysRevB.76.165108} {\bibfield  {journal} {\bibinfo  {journal} {Phys. Rev. B}\ }\textbf {\bibinfo {volume} {76}},\ \bibinfo {pages} {165108} (\bibinfo {year} {2007})}\BibitemShut {NoStop}%
\bibitem [{\citenamefont {Caruso}\ \emph {et~al.}(2017)\citenamefont {Caruso}, \citenamefont {Hoesch}, \citenamefont {Achatz}, \citenamefont {Serrano}, \citenamefont {Krisch}, \citenamefont {Bustarret},\ and\ \citenamefont {Giustino}}]{caruso_nonadiabatic_2017}%
  \BibitemOpen
  \bibfield  {author} {\bibinfo {author} {\bibfnamefont {F.}~\bibnamefont {Caruso}}, \bibinfo {author} {\bibfnamefont {M.}~\bibnamefont {Hoesch}}, \bibinfo {author} {\bibfnamefont {P.}~\bibnamefont {Achatz}}, \bibinfo {author} {\bibfnamefont {J.}~\bibnamefont {Serrano}}, \bibinfo {author} {\bibfnamefont {M.}~\bibnamefont {Krisch}}, \bibinfo {author} {\bibfnamefont {E.}~\bibnamefont {Bustarret}},\ and\ \bibinfo {author} {\bibfnamefont {F.}~\bibnamefont {Giustino}},\ }\bibfield  {title} {\bibinfo {title} {Nonadiabatic kohn anomaly in heavily boron-doped diamond},\ }\href {https://doi.org/10.1103/PhysRevLett.119.017001} {\bibfield  {journal} {\bibinfo  {journal} {Phys. Rev. Lett.}\ }\textbf {\bibinfo {volume} {119}},\ \bibinfo {pages} {017001} (\bibinfo {year} {2017})}\BibitemShut {NoStop}%
\bibitem [{\citenamefont {Novko}\ \emph {et~al.}(2020)\citenamefont {Novko}, \citenamefont {Caruso}, \citenamefont {Draxl},\ and\ \citenamefont {Cappelluti}}]{NovkoCaruso2020}%
  \BibitemOpen
  \bibfield  {author} {\bibinfo {author} {\bibfnamefont {D.}~\bibnamefont {Novko}}, \bibinfo {author} {\bibfnamefont {F.}~\bibnamefont {Caruso}}, \bibinfo {author} {\bibfnamefont {C.}~\bibnamefont {Draxl}},\ and\ \bibinfo {author} {\bibfnamefont {E.}~\bibnamefont {Cappelluti}},\ }\bibfield  {title} {\bibinfo {title} {Ultrafast hot phonon dynamics in {MgB$_2$} driven by anisotropic electron-phonon coupling},\ }\href {https://doi.org/10.1103/PhysRevLett.124.077001} {\bibfield  {journal} {\bibinfo  {journal} {Phys. Rev. Lett.}\ }\textbf {\bibinfo {volume} {124}},\ \bibinfo {pages} {077001} (\bibinfo {year} {2020})}\BibitemShut {NoStop}%
\bibitem [{\citenamefont {Marini}(2024)}]{Marini2024}%
  \BibitemOpen
  \bibfield  {author} {\bibinfo {author} {\bibfnamefont {A.}~\bibnamefont {Marini}},\ }\bibfield  {title} {\bibinfo {title} {Nonadiabatic effects lead to the breakdown of the semiclassical phonon picture},\ }\href {https://doi.org/10.1103/PhysRevB.110.024306} {\bibfield  {journal} {\bibinfo  {journal} {Phys. Rev. B}\ }\textbf {\bibinfo {volume} {110}},\ \bibinfo {pages} {024306} (\bibinfo {year} {2024})}\BibitemShut {NoStop}%
\bibitem [{\citenamefont {Cheng}\ \emph {et~al.}(2018)\citenamefont {Cheng}, \citenamefont {Teitelbaum}, \citenamefont {Gao},\ and\ \citenamefont {Nelson}}]{Cheng2018}%
  \BibitemOpen
  \bibfield  {author} {\bibinfo {author} {\bibfnamefont {Y.-H.}\ \bibnamefont {Cheng}}, \bibinfo {author} {\bibfnamefont {S.~W.}\ \bibnamefont {Teitelbaum}}, \bibinfo {author} {\bibfnamefont {F.~Y.}\ \bibnamefont {Gao}},\ and\ \bibinfo {author} {\bibfnamefont {K.~A.}\ \bibnamefont {Nelson}},\ }\bibfield  {title} {\bibinfo {title} {Femtosecond laser amorphization of tellurium},\ }\href {https://doi.org/10.1103/PhysRevB.98.134112} {\bibfield  {journal} {\bibinfo  {journal} {Phys. Rev. B}\ }\textbf {\bibinfo {volume} {98}},\ \bibinfo {pages} {134112} (\bibinfo {year} {2018})}\BibitemShut {NoStop}%
\bibitem [{\citenamefont {Stefanucci}\ \emph {et~al.}(2023)\citenamefont {Stefanucci}, \citenamefont {van Leeuwen},\ and\ \citenamefont {Perfetto}}]{Stefanucci2023}%
  \BibitemOpen
  \bibfield  {author} {\bibinfo {author} {\bibfnamefont {G.}~\bibnamefont {Stefanucci}}, \bibinfo {author} {\bibfnamefont {R.}~\bibnamefont {van Leeuwen}},\ and\ \bibinfo {author} {\bibfnamefont {E.}~\bibnamefont {Perfetto}},\ }\bibfield  {title} {\bibinfo {title} {In and out-of-equilibrium ab initio theory of electrons and phonons},\ }\href {https://doi.org/10.1103/PhysRevX.13.031026} {\bibfield  {journal} {\bibinfo  {journal} {Phys. Rev. X}\ }\textbf {\bibinfo {volume} {13}},\ \bibinfo {pages} {031026} (\bibinfo {year} {2023})}\BibitemShut {NoStop}%
\bibitem [{\citenamefont {van Leeuwen}(2004)}]{van_leeuwen2004}%
  \BibitemOpen
  \bibfield  {author} {\bibinfo {author} {\bibfnamefont {R.}~\bibnamefont {van Leeuwen}},\ }\bibfield  {title} {\bibinfo {title} {First-principles approach to the electron-phonon interaction},\ }\href {https://doi.org/10.1103/PhysRevB.69.115110} {\bibfield  {journal} {\bibinfo  {journal} {Phys. Rev. B}\ }\textbf {\bibinfo {volume} {69}},\ \bibinfo {pages} {115110} (\bibinfo {year} {2004})}\BibitemShut {NoStop}%
\bibitem [{\citenamefont {Baroni}\ \emph {et~al.}(2001)\citenamefont {Baroni}, \citenamefont {de~Gironcoli}, \citenamefont {Dal~Corso},\ and\ \citenamefont {Giannozzi}}]{Baroni2001}%
  \BibitemOpen
  \bibfield  {author} {\bibinfo {author} {\bibfnamefont {S.}~\bibnamefont {Baroni}}, \bibinfo {author} {\bibfnamefont {S.}~\bibnamefont {de~Gironcoli}}, \bibinfo {author} {\bibfnamefont {A.}~\bibnamefont {Dal~Corso}},\ and\ \bibinfo {author} {\bibfnamefont {P.}~\bibnamefont {Giannozzi}},\ }\bibfield  {title} {\bibinfo {title} {Phonons and related crystal properties from density-functional perturbation theory},\ }\href {https://doi.org/10.1103/RevModPhys.73.515} {\bibfield  {journal} {\bibinfo  {journal} {Rev. Mod. Phys.}\ }\textbf {\bibinfo {volume} {73}},\ \bibinfo {pages} {515} (\bibinfo {year} {2001})}\BibitemShut {NoStop}%
\bibitem [{\citenamefont {Giannozzi}\ \emph {et~al.}(1991)\citenamefont {Giannozzi}, \citenamefont {de~Gironcoli}, \citenamefont {Pavone},\ and\ \citenamefont {Baroni}}]{Giannozzi1991}%
  \BibitemOpen
  \bibfield  {author} {\bibinfo {author} {\bibfnamefont {P.}~\bibnamefont {Giannozzi}}, \bibinfo {author} {\bibfnamefont {S.}~\bibnamefont {de~Gironcoli}}, \bibinfo {author} {\bibfnamefont {P.}~\bibnamefont {Pavone}},\ and\ \bibinfo {author} {\bibfnamefont {S.}~\bibnamefont {Baroni}},\ }\bibfield  {title} {\bibinfo {title} {Ab initio calculation of phonon dispersions in semiconductors},\ }\href {https://doi.org/10.1103/PhysRevB.43.7231} {\bibfield  {journal} {\bibinfo  {journal} {Phys. Rev. B}\ }\textbf {\bibinfo {volume} {43}},\ \bibinfo {pages} {7231} (\bibinfo {year} {1991})}\BibitemShut {NoStop}%
\bibitem [{SI()}]{SI}%
  \BibitemOpen
  \href@noop {} {}\bibinfo {note} {See Supplemental Material at [URL], Ref.~\cite{ONCV,GGA_PBE,mahan2013many,Allen1972,JohnsonBeaud2009} are included in SI.}\BibitemShut {Stop}%
\bibitem [{\citenamefont {Giustino}(2017)}]{Giustino_RMP2017}%
  \BibitemOpen
  \bibfield  {author} {\bibinfo {author} {\bibfnamefont {F.}~\bibnamefont {Giustino}},\ }\bibfield  {title} {\bibinfo {title} {Electron-phonon interactions from first principles},\ }\href {https://doi.org/10.1103/RevModPhys.89.015003} {\bibfield  {journal} {\bibinfo  {journal} {Rev. Mod. Phys.}\ }\textbf {\bibinfo {volume} {89}},\ \bibinfo {pages} {015003} (\bibinfo {year} {2017})}\BibitemShut {NoStop}%
\bibitem [{\citenamefont {Berges}\ \emph {et~al.}(2023)\citenamefont {Berges}, \citenamefont {Girotto}, \citenamefont {Wehling}, \citenamefont {Marzari},\ and\ \citenamefont {Ponc\'e}}]{BergesGirotto2023}%
  \BibitemOpen
  \bibfield  {author} {\bibinfo {author} {\bibfnamefont {J.}~\bibnamefont {Berges}}, \bibinfo {author} {\bibfnamefont {N.}~\bibnamefont {Girotto}}, \bibinfo {author} {\bibfnamefont {T.}~\bibnamefont {Wehling}}, \bibinfo {author} {\bibfnamefont {N.}~\bibnamefont {Marzari}},\ and\ \bibinfo {author} {\bibfnamefont {S.}~\bibnamefont {Ponc\'e}},\ }\bibfield  {title} {\bibinfo {title} {Phonon self-energy corrections: To screen, or not to screen},\ }\href {https://doi.org/10.1103/PhysRevX.13.041009} {\bibfield  {journal} {\bibinfo  {journal} {Phys. Rev. X}\ }\textbf {\bibinfo {volume} {13}},\ \bibinfo {pages} {041009} (\bibinfo {year} {2023})}\BibitemShut {NoStop}%
\bibitem [{\citenamefont {Marini}(2023)}]{Marini2023}%
  \BibitemOpen
  \bibfield  {author} {\bibinfo {author} {\bibfnamefont {A.}~\bibnamefont {Marini}},\ }\bibfield  {title} {\bibinfo {title} {Equilibrium and out-of-equilibrium realistic phonon self-energy free from overscreening},\ }\href {https://doi.org/10.1103/PhysRevB.107.024305} {\bibfield  {journal} {\bibinfo  {journal} {Phys. Rev. B}\ }\textbf {\bibinfo {volume} {107}},\ \bibinfo {pages} {024305} (\bibinfo {year} {2023})}\BibitemShut {NoStop}%
\bibitem [{\citenamefont {{Caldarelli}}\ \emph {et~al.}(2024)\citenamefont {{Caldarelli}}, \citenamefont {{Guandalini}}, \citenamefont {{Macheda}},\ and\ \citenamefont {{Mauri}}}]{Caldarelli2024}%
  \BibitemOpen
  \bibfield  {author} {\bibinfo {author} {\bibfnamefont {G.}~\bibnamefont {{Caldarelli}}}, \bibinfo {author} {\bibfnamefont {A.}~\bibnamefont {{Guandalini}}}, \bibinfo {author} {\bibfnamefont {F.}~\bibnamefont {{Macheda}}},\ and\ \bibinfo {author} {\bibfnamefont {F.}~\bibnamefont {{Mauri}}},\ }\bibfield  {title} {\bibinfo {title} {{Variational formulation of dynamical electronic response functions in presence of nonlocal exchange interactions}},\ }\href {https://doi.org/10.48550/arXiv.2410.22889} {\bibfield  {journal} {\bibinfo  {journal} {arXiv e-prints}\ ,\ \bibinfo {pages} {arXiv:2410.22889}} (\bibinfo {year} {2024})}\BibitemShut {NoStop}%
\bibitem [{\citenamefont {Stefanucci}\ and\ \citenamefont {Perfetto}(2025)}]{StefanucciPerfetto2025}%
  \BibitemOpen
  \bibfield  {author} {\bibinfo {author} {\bibfnamefont {G.}~\bibnamefont {Stefanucci}}\ and\ \bibinfo {author} {\bibfnamefont {E.}~\bibnamefont {Perfetto}},\ }\bibfield  {title} {\bibinfo {title} {Exact formula with two dynamically screened electron-phonon couplings for positive phonon-linewidths approximations},\ }\href {https://doi.org/10.1103/PhysRevB.111.024307} {\bibfield  {journal} {\bibinfo  {journal} {Phys. Rev. B}\ }\textbf {\bibinfo {volume} {111}},\ \bibinfo {pages} {024307} (\bibinfo {year} {2025})}\BibitemShut {NoStop}%
\bibitem [{\citenamefont {Sun}\ \emph {et~al.}(2021)\citenamefont {Sun}, \citenamefont {Kumar}, \citenamefont {Liu}, \citenamefont {Choi}, \citenamefont {Fang}, \citenamefont {Roesch}, \citenamefont {Tran}, \citenamefont {Casara}, \citenamefont {Priego}, \citenamefont {Chang}, \citenamefont {Moody}, \citenamefont {Silverman}, \citenamefont {Lorenz}, \citenamefont {Scheibner}, \citenamefont {Luo},\ and\ \citenamefont {Li}}]{Sun2021}%
  \BibitemOpen
  \bibfield  {author} {\bibinfo {author} {\bibfnamefont {L.}~\bibnamefont {Sun}}, \bibinfo {author} {\bibfnamefont {P.}~\bibnamefont {Kumar}}, \bibinfo {author} {\bibfnamefont {Z.}~\bibnamefont {Liu}}, \bibinfo {author} {\bibfnamefont {J.}~\bibnamefont {Choi}}, \bibinfo {author} {\bibfnamefont {B.}~\bibnamefont {Fang}}, \bibinfo {author} {\bibfnamefont {S.}~\bibnamefont {Roesch}}, \bibinfo {author} {\bibfnamefont {K.}~\bibnamefont {Tran}}, \bibinfo {author} {\bibfnamefont {J.}~\bibnamefont {Casara}}, \bibinfo {author} {\bibfnamefont {E.}~\bibnamefont {Priego}}, \bibinfo {author} {\bibfnamefont {Y.-M.}\ \bibnamefont {Chang}}, \bibinfo {author} {\bibfnamefont {G.}~\bibnamefont {Moody}}, \bibinfo {author} {\bibfnamefont {K.~L.}\ \bibnamefont {Silverman}}, \bibinfo {author} {\bibfnamefont {V.~O.}\ \bibnamefont {Lorenz}}, \bibinfo {author} {\bibfnamefont {M.}~\bibnamefont {Scheibner}}, \bibinfo {author} {\bibfnamefont {T.}~\bibnamefont {Luo}},\ and\ \bibinfo {author} {\bibfnamefont {X.}~\bibnamefont {Li}},\
  }\bibfield  {title} {\bibinfo {title} {Phonon dephasing dynamics in {MoS}$_2$},\ }\href {https://doi.org/10.1021/acs.nanolett.0c04368} {\bibfield  {journal} {\bibinfo  {journal} {Nano Lett.}\ }\textbf {\bibinfo {volume} {21}},\ \bibinfo {pages} {1434} (\bibinfo {year} {2021})}\BibitemShut {NoStop}%
\bibitem [{\citenamefont {Sayers}\ \emph {et~al.}(2023)\citenamefont {Sayers}, \citenamefont {Genco}, \citenamefont {Trovatello}, \citenamefont {Conte}, \citenamefont {Khaustov}, \citenamefont {Cervantes-Villanueva}, \citenamefont {Sangalli}, \citenamefont {Molina-Sanchez}, \citenamefont {Coletti}, \citenamefont {Gadermaier},\ and\ \citenamefont {Cerullo}}]{Sayers2023}%
  \BibitemOpen
  \bibfield  {author} {\bibinfo {author} {\bibfnamefont {C.~J.}\ \bibnamefont {Sayers}}, \bibinfo {author} {\bibfnamefont {A.}~\bibnamefont {Genco}}, \bibinfo {author} {\bibfnamefont {C.}~\bibnamefont {Trovatello}}, \bibinfo {author} {\bibfnamefont {S.~D.}\ \bibnamefont {Conte}}, \bibinfo {author} {\bibfnamefont {V.~O.}\ \bibnamefont {Khaustov}}, \bibinfo {author} {\bibfnamefont {J.}~\bibnamefont {Cervantes-Villanueva}}, \bibinfo {author} {\bibfnamefont {D.}~\bibnamefont {Sangalli}}, \bibinfo {author} {\bibfnamefont {A.}~\bibnamefont {Molina-Sanchez}}, \bibinfo {author} {\bibfnamefont {C.}~\bibnamefont {Coletti}}, \bibinfo {author} {\bibfnamefont {C.}~\bibnamefont {Gadermaier}},\ and\ \bibinfo {author} {\bibfnamefont {G.}~\bibnamefont {Cerullo}},\ }\bibfield  {title} {\bibinfo {title} {Strong coupling of coherent phonons to excitons in semiconducting monolayer {MoTe$_2$}},\ }\href {https://doi.org/10.1021/acs.nanolett.3c01936} {\bibfield  {journal} {\bibinfo  {journal} {Nano Lett.}\ }\textbf {\bibinfo
  {volume} {23}},\ \bibinfo {pages} {9235} (\bibinfo {year} {2023})}\BibitemShut {NoStop}%
\bibitem [{\citenamefont {Trovatello}\ \emph {et~al.}(2020)\citenamefont {Trovatello}, \citenamefont {Miranda}, \citenamefont {Molina-Sánchez}, \citenamefont {Borrego-Varillas}, \citenamefont {Manzoni}, \citenamefont {Moretti}, \citenamefont {Ganzer}, \citenamefont {Maiuri}, \citenamefont {Wang}, \citenamefont {Dumcenco}, \citenamefont {Kis}, \citenamefont {Wirtz}, \citenamefont {Marini}, \citenamefont {Soavi}, \citenamefont {Ferrari}, \citenamefont {Cerullo}, \citenamefont {Sangalli},\ and\ \citenamefont {Conte}}]{Trovatello2020}%
  \BibitemOpen
  \bibfield  {author} {\bibinfo {author} {\bibfnamefont {C.}~\bibnamefont {Trovatello}}, \bibinfo {author} {\bibfnamefont {H.~P.~C.}\ \bibnamefont {Miranda}}, \bibinfo {author} {\bibfnamefont {A.}~\bibnamefont {Molina-Sánchez}}, \bibinfo {author} {\bibfnamefont {R.}~\bibnamefont {Borrego-Varillas}}, \bibinfo {author} {\bibfnamefont {C.}~\bibnamefont {Manzoni}}, \bibinfo {author} {\bibfnamefont {L.}~\bibnamefont {Moretti}}, \bibinfo {author} {\bibfnamefont {L.}~\bibnamefont {Ganzer}}, \bibinfo {author} {\bibfnamefont {M.}~\bibnamefont {Maiuri}}, \bibinfo {author} {\bibfnamefont {J.}~\bibnamefont {Wang}}, \bibinfo {author} {\bibfnamefont {D.}~\bibnamefont {Dumcenco}}, \bibinfo {author} {\bibfnamefont {A.}~\bibnamefont {Kis}}, \bibinfo {author} {\bibfnamefont {L.}~\bibnamefont {Wirtz}}, \bibinfo {author} {\bibfnamefont {A.}~\bibnamefont {Marini}}, \bibinfo {author} {\bibfnamefont {G.}~\bibnamefont {Soavi}}, \bibinfo {author} {\bibfnamefont {A.~C.}\ \bibnamefont {Ferrari}}, \bibinfo {author} {\bibfnamefont
  {G.}~\bibnamefont {Cerullo}}, \bibinfo {author} {\bibfnamefont {D.}~\bibnamefont {Sangalli}},\ and\ \bibinfo {author} {\bibfnamefont {S.~D.}\ \bibnamefont {Conte}},\ }\bibfield  {title} {\bibinfo {title} {Strongly coupled coherent phonons in single-layer {MoS$_2$}},\ }\href {https://doi.org/10.1021/acsnano.0c00309} {\bibfield  {journal} {\bibinfo  {journal} {ACS Nano}\ }\textbf {\bibinfo {volume} {14}},\ \bibinfo {pages} {5700} (\bibinfo {year} {2020})}\BibitemShut {NoStop}%
\bibitem [{\citenamefont {Jeong}\ \emph {et~al.}(2016)\citenamefont {Jeong}, \citenamefont {Jin}, \citenamefont {Rhim}, \citenamefont {Debbichi}, \citenamefont {Park}, \citenamefont {Jang}, \citenamefont {Lee}, \citenamefont {Chae}, \citenamefont {Lee}, \citenamefont {Kim}, \citenamefont {Jung},\ and\ \citenamefont {Yee}}]{Jeong2016}%
  \BibitemOpen
  \bibfield  {author} {\bibinfo {author} {\bibfnamefont {T.~Y.}\ \bibnamefont {Jeong}}, \bibinfo {author} {\bibfnamefont {B.~M.}\ \bibnamefont {Jin}}, \bibinfo {author} {\bibfnamefont {S.~H.}\ \bibnamefont {Rhim}}, \bibinfo {author} {\bibfnamefont {L.}~\bibnamefont {Debbichi}}, \bibinfo {author} {\bibfnamefont {J.}~\bibnamefont {Park}}, \bibinfo {author} {\bibfnamefont {Y.~D.}\ \bibnamefont {Jang}}, \bibinfo {author} {\bibfnamefont {H.~R.}\ \bibnamefont {Lee}}, \bibinfo {author} {\bibfnamefont {D.-H.}\ \bibnamefont {Chae}}, \bibinfo {author} {\bibfnamefont {D.}~\bibnamefont {Lee}}, \bibinfo {author} {\bibfnamefont {Y.-H.}\ \bibnamefont {Kim}}, \bibinfo {author} {\bibfnamefont {S.}~\bibnamefont {Jung}},\ and\ \bibinfo {author} {\bibfnamefont {K.~J.}\ \bibnamefont {Yee}},\ }\bibfield  {title} {\bibinfo {title} {Coherent lattice vibrations in mono- and few-layer {WSe$_2$}},\ }\href {https://doi.org/10.1021/acsnano.6b02253} {\bibfield  {journal} {\bibinfo  {journal} {ACS Nano}\ }\textbf {\bibinfo {volume} {10}},\
  \bibinfo {pages} {5560} (\bibinfo {year} {2016})}\BibitemShut {NoStop}%
\bibitem [{\citenamefont {Li}\ \emph {et~al.}(2014)\citenamefont {Li}, \citenamefont {Carrete}, \citenamefont {{A. Katcho}},\ and\ \citenamefont {Mingo}}]{LI20141747}%
  \BibitemOpen
  \bibfield  {author} {\bibinfo {author} {\bibfnamefont {W.}~\bibnamefont {Li}}, \bibinfo {author} {\bibfnamefont {J.}~\bibnamefont {Carrete}}, \bibinfo {author} {\bibfnamefont {N.}~\bibnamefont {{A. Katcho}}},\ and\ \bibinfo {author} {\bibfnamefont {N.}~\bibnamefont {Mingo}},\ }\bibfield  {title} {\bibinfo {title} {{ShengBTE}: A solver of the boltzmann transport equation for phonons},\ }\href {https://doi.org/https://doi.org/10.1016/j.cpc.2014.02.015} {\bibfield  {journal} {\bibinfo  {journal} {Comput. Phys. Commun.}\ }\textbf {\bibinfo {volume} {185}},\ \bibinfo {pages} {1747} (\bibinfo {year} {2014})}\BibitemShut {NoStop}%
\bibitem [{\citenamefont {Lazzeri}\ \emph {et~al.}(2003)\citenamefont {Lazzeri}, \citenamefont {Calandra},\ and\ \citenamefont {Mauri}}]{Lazzeri2003}%
  \BibitemOpen
  \bibfield  {author} {\bibinfo {author} {\bibfnamefont {M.}~\bibnamefont {Lazzeri}}, \bibinfo {author} {\bibfnamefont {M.}~\bibnamefont {Calandra}},\ and\ \bibinfo {author} {\bibfnamefont {F.}~\bibnamefont {Mauri}},\ }\bibfield  {title} {\bibinfo {title} {Anharmonic phonon frequency shift in {MgB$_2$}},\ }\href {https://doi.org/10.1103/PhysRevB.68.220509} {\bibfield  {journal} {\bibinfo  {journal} {Phys. Rev. B}\ }\textbf {\bibinfo {volume} {68}},\ \bibinfo {pages} {220509} (\bibinfo {year} {2003})}\BibitemShut {NoStop}%
\bibitem [{\citenamefont {Cowley}(1968)}]{Cowley1968}%
  \BibitemOpen
  \bibfield  {author} {\bibinfo {author} {\bibfnamefont {R.~A.}\ \bibnamefont {Cowley}},\ }\bibfield  {title} {\bibinfo {title} {Anharmonic crystals},\ }\href {https://doi.org/10.1088/0034-4885/31/1/303} {\bibfield  {journal} {\bibinfo  {journal} {Rep. Prog. Phys.}\ }\textbf {\bibinfo {volume} {31}},\ \bibinfo {pages} {123} (\bibinfo {year} {1968})}\BibitemShut {NoStop}%
\bibitem [{\citenamefont {Thompson}(1963)}]{Thompson1963}%
  \BibitemOpen
  \bibfield  {author} {\bibinfo {author} {\bibfnamefont {B.~V.}\ \bibnamefont {Thompson}},\ }\bibfield  {title} {\bibinfo {title} {Neutron scattering by an anharmonic crystal},\ }\href {https://doi.org/10.1103/PhysRev.131.1420} {\bibfield  {journal} {\bibinfo  {journal} {Phys. Rev.}\ }\textbf {\bibinfo {volume} {131}},\ \bibinfo {pages} {1420} (\bibinfo {year} {1963})}\BibitemShut {NoStop}%
\bibitem [{\citenamefont {Maradudin}\ and\ \citenamefont {Fein}(1962)}]{Maradudin_Fein1962}%
  \BibitemOpen
  \bibfield  {author} {\bibinfo {author} {\bibfnamefont {A.~A.}\ \bibnamefont {Maradudin}}\ and\ \bibinfo {author} {\bibfnamefont {A.~E.}\ \bibnamefont {Fein}},\ }\bibfield  {title} {\bibinfo {title} {Scattering of neutrons by an anharmonic crystal},\ }\href {https://doi.org/10.1103/PhysRev.128.2589} {\bibfield  {journal} {\bibinfo  {journal} {Phys. Rev.}\ }\textbf {\bibinfo {volume} {128}},\ \bibinfo {pages} {2589} (\bibinfo {year} {1962})}\BibitemShut {NoStop}%
\bibitem [{\citenamefont {Lax}(1964)}]{LAX1964487}%
  \BibitemOpen
  \bibfield  {author} {\bibinfo {author} {\bibfnamefont {M.}~\bibnamefont {Lax}},\ }\bibfield  {title} {\bibinfo {title} {Quantum relaxation, the shape of lattice absorption and inelastic neutron scattering lines},\ }\href {https://doi.org/https://doi.org/10.1016/0022-3697(64)90122-2} {\bibfield  {journal} {\bibinfo  {journal} {J. Phys. Chem. Solids}\ }\textbf {\bibinfo {volume} {25}},\ \bibinfo {pages} {487} (\bibinfo {year} {1964})}\BibitemShut {NoStop}%
\bibitem [{\citenamefont {Hase}\ \emph {et~al.}(1998)\citenamefont {Hase}, \citenamefont {Mizoguchi}, \citenamefont {Harima}, \citenamefont {Nakashima},\ and\ \citenamefont {Sakai}}]{Hase1998}%
  \BibitemOpen
  \bibfield  {author} {\bibinfo {author} {\bibfnamefont {M.}~\bibnamefont {Hase}}, \bibinfo {author} {\bibfnamefont {K.}~\bibnamefont {Mizoguchi}}, \bibinfo {author} {\bibfnamefont {H.}~\bibnamefont {Harima}}, \bibinfo {author} {\bibfnamefont {S.-i.}\ \bibnamefont {Nakashima}},\ and\ \bibinfo {author} {\bibfnamefont {K.}~\bibnamefont {Sakai}},\ }\bibfield  {title} {\bibinfo {title} {Dynamics of coherent phonons in bismuth generated by ultrashort laser pulses},\ }\href {https://doi.org/10.1103/PhysRevB.58.5448} {\bibfield  {journal} {\bibinfo  {journal} {Phys. Rev. B}\ }\textbf {\bibinfo {volume} {58}},\ \bibinfo {pages} {5448} (\bibinfo {year} {1998})}\BibitemShut {NoStop}%
\bibitem [{\citenamefont {Hase}\ \emph {et~al.}(2015)\citenamefont {Hase}, \citenamefont {Ushida},\ and\ \citenamefont {Kitajima}}]{Hase2015}%
  \BibitemOpen
  \bibfield  {author} {\bibinfo {author} {\bibfnamefont {M.}~\bibnamefont {Hase}}, \bibinfo {author} {\bibfnamefont {K.}~\bibnamefont {Ushida}},\ and\ \bibinfo {author} {\bibfnamefont {M.}~\bibnamefont {Kitajima}},\ }\bibfield  {title} {\bibinfo {title} {Anharmonic decay of coherent optical phonons in antimony},\ }\href {https://doi.org/10.7566/JPSJ.84.024708} {\bibfield  {journal} {\bibinfo  {journal} {J. Phys. Soc. Jpn.}\ }\textbf {\bibinfo {volume} {84}},\ \bibinfo {pages} {024708} (\bibinfo {year} {2015})}\BibitemShut {NoStop}%
\bibitem [{\citenamefont {Ishioka}\ and\ \citenamefont {Misochko}(2024)}]{Ishioka2024}%
  \BibitemOpen
  \bibfield  {author} {\bibinfo {author} {\bibfnamefont {K.}~\bibnamefont {Ishioka}}\ and\ \bibinfo {author} {\bibfnamefont {O.~V.}\ \bibnamefont {Misochko}},\ }\bibfield  {title} {\bibinfo {title} {Suppression of shear ionic motions in bismuth by coupling with large-amplitude internal displacement},\ }\href {https://doi.org/10.1103/PhysRevB.110.094313} {\bibfield  {journal} {\bibinfo  {journal} {Phys. Rev. B}\ }\textbf {\bibinfo {volume} {110}},\ \bibinfo {pages} {094313} (\bibinfo {year} {2024})}\BibitemShut {NoStop}%
\bibitem [{\citenamefont {{Emeis}}\ \emph {et~al.}(2024)\citenamefont {{Emeis}}, \citenamefont {{Jauernik}}, \citenamefont {{Dahiya}}, \citenamefont {{Pan}}, \citenamefont {{Jensen}}, \citenamefont {{Hein}}, \citenamefont {{Bauer}},\ and\ \citenamefont {{Caruso}}}]{EmeisJauernik2024}%
  \BibitemOpen
  \bibfield  {author} {\bibinfo {author} {\bibfnamefont {C.}~\bibnamefont {{Emeis}}}, \bibinfo {author} {\bibfnamefont {S.}~\bibnamefont {{Jauernik}}}, \bibinfo {author} {\bibfnamefont {S.}~\bibnamefont {{Dahiya}}}, \bibinfo {author} {\bibfnamefont {Y.}~\bibnamefont {{Pan}}}, \bibinfo {author} {\bibfnamefont {C.~E.}\ \bibnamefont {{Jensen}}}, \bibinfo {author} {\bibfnamefont {P.}~\bibnamefont {{Hein}}}, \bibinfo {author} {\bibfnamefont {M.}~\bibnamefont {{Bauer}}},\ and\ \bibinfo {author} {\bibfnamefont {F.}~\bibnamefont {{Caruso}}},\ }\bibfield  {title} {\bibinfo {title} {{Coherent Phonons and Quasiparticle Renormalization in Semimetals from First Principles}},\ }\href {https://doi.org/10.48550/arXiv.2407.17118} {\bibfield  {journal} {\bibinfo  {journal} {arXiv e-prints}\ ,\ \bibinfo {pages} {arXiv:2407.17118}} (\bibinfo {year} {2024})}\BibitemShut {NoStop}%
\bibitem [{\citenamefont {Ponc\'{e}}\ \emph {et~al.}(2016)\citenamefont {Ponc\'{e}}, \citenamefont {Margine}, \citenamefont {Verdi},\ and\ \citenamefont {Giustino}}]{bib:epw}%
  \BibitemOpen
  \bibfield  {author} {\bibinfo {author} {\bibfnamefont {S.}~\bibnamefont {Ponc\'{e}}}, \bibinfo {author} {\bibfnamefont {E.}~\bibnamefont {Margine}}, \bibinfo {author} {\bibfnamefont {C.}~\bibnamefont {Verdi}},\ and\ \bibinfo {author} {\bibfnamefont {F.}~\bibnamefont {Giustino}},\ }\bibfield  {title} {\bibinfo {title} {Epw: Electron–phonon coupling, transport and superconducting properties using maximally localized wannier functions},\ }\href {https://doi.org/https://doi.org/10.1016/j.cpc.2016.07.028} {\bibfield  {journal} {\bibinfo  {journal} {Comp. Phys. Commun.}\ }\textbf {\bibinfo {volume} {209}},\ \bibinfo {pages} {116} (\bibinfo {year} {2016})}\BibitemShut {NoStop}%
\bibitem [{\citenamefont {Giannozzi}\ \emph {et~al.}(2017)\citenamefont {Giannozzi}, \citenamefont {Andreussi}, \citenamefont {Brumme}, \citenamefont {Bunau}, \citenamefont {Nardelli}, \citenamefont {Calandra}, \citenamefont {Car}, \citenamefont {Cavazzoni}, \citenamefont {Ceresoli}, \citenamefont {Cococcioni} \emph {et~al.}}]{Giannozzi2017}%
  \BibitemOpen
  \bibfield  {author} {\bibinfo {author} {\bibfnamefont {P.}~\bibnamefont {Giannozzi}}, \bibinfo {author} {\bibfnamefont {O.}~\bibnamefont {Andreussi}}, \bibinfo {author} {\bibfnamefont {T.}~\bibnamefont {Brumme}}, \bibinfo {author} {\bibfnamefont {O.}~\bibnamefont {Bunau}}, \bibinfo {author} {\bibfnamefont {M.~B.}\ \bibnamefont {Nardelli}}, \bibinfo {author} {\bibfnamefont {M.}~\bibnamefont {Calandra}}, \bibinfo {author} {\bibfnamefont {R.}~\bibnamefont {Car}}, \bibinfo {author} {\bibfnamefont {C.}~\bibnamefont {Cavazzoni}}, \bibinfo {author} {\bibfnamefont {D.}~\bibnamefont {Ceresoli}}, \bibinfo {author} {\bibfnamefont {M.}~\bibnamefont {Cococcioni}}, \emph {et~al.},\ }\bibfield  {title} {\bibinfo {title} {Advanced capabilities for materials modelling with quantum {ESPRESSO}},\ }\href {https://doi.org/10.1088/1361-648x/aa8f79} {\bibfield  {journal} {\bibinfo  {journal} {J. Phys.: Condens. Matter}\ }\textbf {\bibinfo {volume} {29}},\ \bibinfo {pages} {465901} (\bibinfo {year} {2017})}\BibitemShut {NoStop}%
\bibitem [{\citenamefont {Pizzi}\ \emph {et~al.}(2020)\citenamefont {Pizzi}, \citenamefont {Vitale}, \citenamefont {Arita}, \citenamefont {Blügel}, \citenamefont {Freimuth}, \citenamefont {Géranton}, \citenamefont {Gibertini}, \citenamefont {Gresch}, \citenamefont {Johnson}, \citenamefont {Koretsune}, \citenamefont {Ibañez-Azpiroz}, \citenamefont {Lee}, \citenamefont {Lihm}, \citenamefont {Marchand}, \citenamefont {Marrazzo}, \citenamefont {Mokrousov}, \citenamefont {Mustafa}, \citenamefont {Nohara}, \citenamefont {Nomura}, \citenamefont {Paulatto}, \citenamefont {Poncé}, \citenamefont {Ponweiser}, \citenamefont {Qiao}, \citenamefont {Thöle}, \citenamefont {Tsirkin}, \citenamefont {Wierzbowska}, \citenamefont {Marzari}, \citenamefont {Vanderbilt}, \citenamefont {Souza}, \citenamefont {Mostofi},\ and\ \citenamefont {Yates}}]{pizzi2020wannier90}%
  \BibitemOpen
  \bibfield  {author} {\bibinfo {author} {\bibfnamefont {G.}~\bibnamefont {Pizzi}}, \bibinfo {author} {\bibfnamefont {V.}~\bibnamefont {Vitale}}, \bibinfo {author} {\bibfnamefont {R.}~\bibnamefont {Arita}}, \bibinfo {author} {\bibfnamefont {S.}~\bibnamefont {Blügel}}, \bibinfo {author} {\bibfnamefont {F.}~\bibnamefont {Freimuth}}, \bibinfo {author} {\bibfnamefont {G.}~\bibnamefont {Géranton}}, \bibinfo {author} {\bibfnamefont {M.}~\bibnamefont {Gibertini}}, \bibinfo {author} {\bibfnamefont {D.}~\bibnamefont {Gresch}}, \bibinfo {author} {\bibfnamefont {C.}~\bibnamefont {Johnson}}, \bibinfo {author} {\bibfnamefont {T.}~\bibnamefont {Koretsune}}, \bibinfo {author} {\bibfnamefont {J.}~\bibnamefont {Ibañez-Azpiroz}}, \bibinfo {author} {\bibfnamefont {H.}~\bibnamefont {Lee}}, \bibinfo {author} {\bibfnamefont {J.-M.}\ \bibnamefont {Lihm}}, \bibinfo {author} {\bibfnamefont {D.}~\bibnamefont {Marchand}}, \bibinfo {author} {\bibfnamefont {A.}~\bibnamefont {Marrazzo}}, \bibinfo {author} {\bibfnamefont
  {Y.}~\bibnamefont {Mokrousov}}, \bibinfo {author} {\bibfnamefont {J.~I.}\ \bibnamefont {Mustafa}}, \bibinfo {author} {\bibfnamefont {Y.}~\bibnamefont {Nohara}}, \bibinfo {author} {\bibfnamefont {Y.}~\bibnamefont {Nomura}}, \bibinfo {author} {\bibfnamefont {L.}~\bibnamefont {Paulatto}}, \bibinfo {author} {\bibfnamefont {S.}~\bibnamefont {Poncé}}, \bibinfo {author} {\bibfnamefont {T.}~\bibnamefont {Ponweiser}}, \bibinfo {author} {\bibfnamefont {J.}~\bibnamefont {Qiao}}, \bibinfo {author} {\bibfnamefont {F.}~\bibnamefont {Thöle}}, \bibinfo {author} {\bibfnamefont {S.~S.}\ \bibnamefont {Tsirkin}}, \bibinfo {author} {\bibfnamefont {M.}~\bibnamefont {Wierzbowska}}, \bibinfo {author} {\bibfnamefont {N.}~\bibnamefont {Marzari}}, \bibinfo {author} {\bibfnamefont {D.}~\bibnamefont {Vanderbilt}}, \bibinfo {author} {\bibfnamefont {I.}~\bibnamefont {Souza}}, \bibinfo {author} {\bibfnamefont {A.~A.}\ \bibnamefont {Mostofi}},\ and\ \bibinfo {author} {\bibfnamefont {J.~R.}\ \bibnamefont {Yates}},\ }\bibfield  {title}
  {\bibinfo {title} {Wannier90 as a community code: new features and applications},\ }\href {https://doi.org/10.1088/1361-648X/ab51ff} {\bibfield  {journal} {\bibinfo  {journal} {J. Phys.: Condens. Matter}\ }\textbf {\bibinfo {volume} {32}},\ \bibinfo {pages} {165902} (\bibinfo {year} {2020})}\BibitemShut {NoStop}%
\bibitem [{\citenamefont {Zeiger}\ \emph {et~al.}(1992)\citenamefont {Zeiger}, \citenamefont {Vidal}, \citenamefont {Cheng}, \citenamefont {Ippen}, \citenamefont {Dresselhaus},\ and\ \citenamefont {Dresselhaus}}]{Zeiger1992}%
  \BibitemOpen
  \bibfield  {author} {\bibinfo {author} {\bibfnamefont {H.~J.}\ \bibnamefont {Zeiger}}, \bibinfo {author} {\bibfnamefont {J.}~\bibnamefont {Vidal}}, \bibinfo {author} {\bibfnamefont {T.~K.}\ \bibnamefont {Cheng}}, \bibinfo {author} {\bibfnamefont {E.~P.}\ \bibnamefont {Ippen}}, \bibinfo {author} {\bibfnamefont {G.}~\bibnamefont {Dresselhaus}},\ and\ \bibinfo {author} {\bibfnamefont {M.~S.}\ \bibnamefont {Dresselhaus}},\ }\bibfield  {title} {\bibinfo {title} {Theory for displacive excitation of coherent phonons},\ }\href {https://doi.org/10.1103/PhysRevB.45.768} {\bibfield  {journal} {\bibinfo  {journal} {Phys. Rev. B}\ }\textbf {\bibinfo {volume} {45}},\ \bibinfo {pages} {768} (\bibinfo {year} {1992})}\BibitemShut {NoStop}%
\bibitem [{\citenamefont {Kuznetsov}\ and\ \citenamefont {Stanton}(1994)}]{Kuznetsov1994}%
  \BibitemOpen
  \bibfield  {author} {\bibinfo {author} {\bibfnamefont {A.~V.}\ \bibnamefont {Kuznetsov}}\ and\ \bibinfo {author} {\bibfnamefont {C.~J.}\ \bibnamefont {Stanton}},\ }\bibfield  {title} {\bibinfo {title} {Theory of coherent phonon oscillations in semiconductors},\ }\href {https://doi.org/10.1103/PhysRevLett.73.3243} {\bibfield  {journal} {\bibinfo  {journal} {Phys. Rev. Lett.}\ }\textbf {\bibinfo {volume} {73}},\ \bibinfo {pages} {3243} (\bibinfo {year} {1994})}\BibitemShut {NoStop}%
\bibitem [{\citenamefont {Li}\ \emph {et~al.}(2013)\citenamefont {Li}, \citenamefont {Chen}, \citenamefont {Reis}, \citenamefont {Fahy},\ and\ \citenamefont {Merlin}}]{LiChenReis2013}%
  \BibitemOpen
  \bibfield  {author} {\bibinfo {author} {\bibfnamefont {J.~J.}\ \bibnamefont {Li}}, \bibinfo {author} {\bibfnamefont {J.}~\bibnamefont {Chen}}, \bibinfo {author} {\bibfnamefont {D.~A.}\ \bibnamefont {Reis}}, \bibinfo {author} {\bibfnamefont {S.}~\bibnamefont {Fahy}},\ and\ \bibinfo {author} {\bibfnamefont {R.}~\bibnamefont {Merlin}},\ }\bibfield  {title} {\bibinfo {title} {Optical probing of ultrafast electronic decay in bi and sb with slow phonons},\ }\href {https://doi.org/10.1103/PhysRevLett.110.047401} {\bibfield  {journal} {\bibinfo  {journal} {Phys. Rev. Lett.}\ }\textbf {\bibinfo {volume} {110}},\ \bibinfo {pages} {047401} (\bibinfo {year} {2013})}\BibitemShut {NoStop}%
\bibitem [{\citenamefont {Novko}(2018)}]{Novko2018}%
  \BibitemOpen
  \bibfield  {author} {\bibinfo {author} {\bibfnamefont {D.}~\bibnamefont {Novko}},\ }\bibfield  {title} {\bibinfo {title} {Nonadiabatic coupling effects in {MgB$_2$} reexamined},\ }\href {https://doi.org/10.1103/PhysRevB.98.041112} {\bibfield  {journal} {\bibinfo  {journal} {Phys. Rev. B}\ }\textbf {\bibinfo {volume} {98}},\ \bibinfo {pages} {041112} (\bibinfo {year} {2018})}\BibitemShut {NoStop}%
\bibitem [{\citenamefont {Lihm}\ \emph {et~al.}(2024)\citenamefont {Lihm}, \citenamefont {Ponc\'e},\ and\ \citenamefont {Park}}]{LihmPonce2024}%
  \BibitemOpen
  \bibfield  {author} {\bibinfo {author} {\bibfnamefont {J.-M.}\ \bibnamefont {Lihm}}, \bibinfo {author} {\bibfnamefont {S.}~\bibnamefont {Ponc\'e}},\ and\ \bibinfo {author} {\bibfnamefont {C.-H.}\ \bibnamefont {Park}},\ }\bibfield  {title} {\bibinfo {title} {Self-consistent electron lifetimes for electron-phonon scattering},\ }\href {https://doi.org/10.1103/PhysRevB.110.L121106} {\bibfield  {journal} {\bibinfo  {journal} {Phys. Rev. B}\ }\textbf {\bibinfo {volume} {110}},\ \bibinfo {pages} {L121106} (\bibinfo {year} {2024})}\BibitemShut {NoStop}%
\bibitem [{\citenamefont {{Park}}(2024)}]{Park2024}%
  \BibitemOpen
  \bibfield  {author} {\bibinfo {author} {\bibfnamefont {C.-H.}\ \bibnamefont {{Park}}},\ }\bibfield  {title} {\bibinfo {title} {{Non-adiabatic phonon self-energy due to electrons with finite linewidths}},\ }\href {https://doi.org/10.48550/arXiv.2411.12221} {\bibfield  {journal} {\bibinfo  {journal} {arXiv e-prints}\ ,\ \bibinfo {eid} {arXiv:2411.12221}} (\bibinfo {year} {2024})}\BibitemShut {NoStop}%
\bibitem [{\citenamefont {Bonini}\ \emph {et~al.}(2007)\citenamefont {Bonini}, \citenamefont {Lazzeri}, \citenamefont {Marzari},\ and\ \citenamefont {Mauri}}]{Bonini2007}%
  \BibitemOpen
  \bibfield  {author} {\bibinfo {author} {\bibfnamefont {N.}~\bibnamefont {Bonini}}, \bibinfo {author} {\bibfnamefont {M.}~\bibnamefont {Lazzeri}}, \bibinfo {author} {\bibfnamefont {N.}~\bibnamefont {Marzari}},\ and\ \bibinfo {author} {\bibfnamefont {F.}~\bibnamefont {Mauri}},\ }\bibfield  {title} {\bibinfo {title} {Phonon anharmonicities in graphite and graphene},\ }\href {https://doi.org/10.1103/PhysRevLett.99.176802} {\bibfield  {journal} {\bibinfo  {journal} {Phys. Rev. Lett.}\ }\textbf {\bibinfo {volume} {99}},\ \bibinfo {pages} {176802} (\bibinfo {year} {2007})}\BibitemShut {NoStop}%
\bibitem [{\citenamefont {Caruso}(2021)}]{Caruso2021}%
  \BibitemOpen
  \bibfield  {author} {\bibinfo {author} {\bibfnamefont {F.}~\bibnamefont {Caruso}},\ }\bibfield  {title} {\bibinfo {title} {Nonequilibrium lattice dynamics in monolayer {MoS$_2$}},\ }\href {https://doi.org/10.1021/acs.jpclett.0c03616} {\bibfield  {journal} {\bibinfo  {journal} {J. Phys. Chem. Lett.}\ }\textbf {\bibinfo {volume} {12}},\ \bibinfo {pages} {1734} (\bibinfo {year} {2021})}\BibitemShut {NoStop}%
\bibitem [{\citenamefont {Caruso}\ and\ \citenamefont {Novko}(2022)}]{TDBE2}%
  \BibitemOpen
  \bibfield  {author} {\bibinfo {author} {\bibfnamefont {F.}~\bibnamefont {Caruso}}\ and\ \bibinfo {author} {\bibfnamefont {D.}~\bibnamefont {Novko}},\ }\bibfield  {title} {\bibinfo {title} {Ultrafast dynamics of electrons and phonons: from the two-temperature model to the time-dependent boltzmann equation},\ }\href {https://doi.org/10.1080/23746149.2022.2095925} {\bibfield  {journal} {\bibinfo  {journal} {Adv. Phys.: X}\ }\textbf {\bibinfo {volume} {7}},\ \bibinfo {pages} {2095925} (\bibinfo {year} {2022})}\BibitemShut {NoStop}%
\bibitem [{\citenamefont {Pan}\ and\ \citenamefont {Caruso}(2023)}]{PanCaruso2023}%
  \BibitemOpen
  \bibfield  {author} {\bibinfo {author} {\bibfnamefont {Y.}~\bibnamefont {Pan}}\ and\ \bibinfo {author} {\bibfnamefont {F.}~\bibnamefont {Caruso}},\ }\bibfield  {title} {\bibinfo {title} {Vibrational dichroism of chiral valley phonons},\ }\href {https://doi.org/10.1021/acs.nanolett.3c01904} {\bibfield  {journal} {\bibinfo  {journal} {Nano Lett.}\ }\textbf {\bibinfo {volume} {23}},\ \bibinfo {pages} {7463} (\bibinfo {year} {2023})}\BibitemShut {NoStop}%
\bibitem [{\citenamefont {Pan}\ and\ \citenamefont {Caruso}(2024)}]{PanCaruso2024}%
  \BibitemOpen
  \bibfield  {author} {\bibinfo {author} {\bibfnamefont {Y.}~\bibnamefont {Pan}}\ and\ \bibinfo {author} {\bibfnamefont {F.}~\bibnamefont {Caruso}},\ }\bibfield  {title} {\bibinfo {title} {Strain-induced activation of chiral-phonon emission in monolayer {WS}$_2$},\ }\href {https://doi.org/10.1038/s41699-024-00479-4} {\bibfield  {journal} {\bibinfo  {journal} {npj 2D Mater. Appl.}\ }\textbf {\bibinfo {volume} {8}},\ \bibinfo {pages} {42} (\bibinfo {year} {2024})}\BibitemShut {NoStop}%
\bibitem [{\citenamefont {Hamann}(2013)}]{ONCV}%
  \BibitemOpen
  \bibfield  {author} {\bibinfo {author} {\bibfnamefont {D.~R.}\ \bibnamefont {Hamann}},\ }\bibfield  {title} {\bibinfo {title} {Optimized norm-conserving vanderbilt pseudopotentials},\ }\href {https://doi.org/10.1103/PhysRevB.88.085117} {\bibfield  {journal} {\bibinfo  {journal} {Phys. Rev. B}\ }\textbf {\bibinfo {volume} {88}},\ \bibinfo {pages} {085117} (\bibinfo {year} {2013})}\BibitemShut {NoStop}%
\bibitem [{\citenamefont {Perdew}\ \emph {et~al.}(1996)\citenamefont {Perdew}, \citenamefont {Burke},\ and\ \citenamefont {Ernzerhof}}]{GGA_PBE}%
  \BibitemOpen
  \bibfield  {author} {\bibinfo {author} {\bibfnamefont {J.~P.}\ \bibnamefont {Perdew}}, \bibinfo {author} {\bibfnamefont {K.}~\bibnamefont {Burke}},\ and\ \bibinfo {author} {\bibfnamefont {M.}~\bibnamefont {Ernzerhof}},\ }\bibfield  {title} {\bibinfo {title} {Generalized gradient approximation made simple},\ }\href {https://doi.org/10.1103/PhysRevLett.77.3865} {\bibfield  {journal} {\bibinfo  {journal} {Phys. Rev. Lett.}\ }\textbf {\bibinfo {volume} {77}},\ \bibinfo {pages} {3865} (\bibinfo {year} {1996})}\BibitemShut {NoStop}%
\bibitem [{\citenamefont {Mahan}(2000)}]{mahan2013many}%
  \BibitemOpen
  \bibfield  {author} {\bibinfo {author} {\bibfnamefont {G.~D.}\ \bibnamefont {Mahan}},\ }\href {https://doi.org/https://doi.org/10.1007/978-1-4757-5714-9} {\emph {\bibinfo {title} {Many-particle {Physics}}}}\ (\bibinfo  {publisher} {Springer New York, NY},\ \bibinfo {year} {2000})\BibitemShut {NoStop}%
\bibitem [{\citenamefont {Allen}(1972)}]{Allen1972}%
  \BibitemOpen
  \bibfield  {author} {\bibinfo {author} {\bibfnamefont {P.~B.}\ \bibnamefont {Allen}},\ }\bibfield  {title} {\bibinfo {title} {Neutron spectroscopy of superconductors},\ }\href {https://doi.org/10.1103/PhysRevB.6.2577} {\bibfield  {journal} {\bibinfo  {journal} {Phys. Rev. B}\ }\textbf {\bibinfo {volume} {6}},\ \bibinfo {pages} {2577} (\bibinfo {year} {1972})}\BibitemShut {NoStop}%
\bibitem [{\citenamefont {Johnson}\ \emph {et~al.}(2009)\citenamefont {Johnson}, \citenamefont {Beaud}, \citenamefont {Vorobeva}, \citenamefont {Milne}, \citenamefont {Murray}, \citenamefont {Fahy},\ and\ \citenamefont {Ingold}}]{JohnsonBeaud2009}%
  \BibitemOpen
  \bibfield  {author} {\bibinfo {author} {\bibfnamefont {S.~L.}\ \bibnamefont {Johnson}}, \bibinfo {author} {\bibfnamefont {P.}~\bibnamefont {Beaud}}, \bibinfo {author} {\bibfnamefont {E.}~\bibnamefont {Vorobeva}}, \bibinfo {author} {\bibfnamefont {C.~J.}\ \bibnamefont {Milne}}, \bibinfo {author} {\bibfnamefont {E.~D.}\ \bibnamefont {Murray}}, \bibinfo {author} {\bibfnamefont {S.}~\bibnamefont {Fahy}},\ and\ \bibinfo {author} {\bibfnamefont {G.}~\bibnamefont {Ingold}},\ }\bibfield  {title} {\bibinfo {title} {Directly observing squeezed phonon states with femtosecond x-ray diffraction},\ }\href {https://doi.org/10.1103/PhysRevLett.102.175503} {\bibfield  {journal} {\bibinfo  {journal} {Phys. Rev. Lett.}\ }\textbf {\bibinfo {volume} {102}},\ \bibinfo {pages} {175503} (\bibinfo {year} {2009})}\BibitemShut {NoStop}%
\end{thebibliography}
%

\end{document}